\documentclass[]{pasj02} 
\usepackage[switch,mathlines]{lineno} % add line number to manuscript

\jyear{2026}
\Received{2026/08/24}%{yyyy/mm/dd}
\Accepted{2026/10/15}%{yyyy/mm/dd}
\begin{document} 

\title{ MIDSSAR: MId Dispersion Spectrograph for Stellar Activity Research }

%%% begin:list of authors
% Do NOT capitalize all letters in "textsc".
\author{
 Fumihde \textsc{Iwamuro},\altaffilmark{1}\altemailmark\orcid{0000-0002-7419-2629} \email{iwamuro@kusastro.kyoto-u.ac.jp} 
 Daisaku \textsc{Nogami},\altaffilmark{1}\orcid{0000-0001-9588-1872}
 Hiroyuki \textsc{Maehara}\altaffilmark{2}$^,$\altaffilmark{3}\orcid{0000-0003-0332-0811}
 Masaru \textsc{Kino}\altaffilmark{3}
 Kai \textsc{Ikuta}\altaffilmark{4}\orcid{0000-0002-5978-057X}
 Akito \textsc{Tajitsu}\altaffilmark{2}\orcid{0000-0001-8813-9338}
 Sanghee \textsc{Lee}\altaffilmark{1}\orcid{0009-0007-7374-5133}
 Miho \textsc{Kawabata}\altaffilmark{3}\orcid{0000-0002-4540-4928}
 Keisuke \textsc{Isogai}\altaffilmark{3}$^,$\altaffilmark{5}\orcid{0000-0002-6480-3799}
 Kenta \textsc{Taguchi}\altaffilmark{3}\orcid{0000-0002-8482-8993}
 and
 Kazunari \textsc{Shibata},\altaffilmark{6}$^,$\altaffilmark{7}\orcid{0000-0003-1206-7889}
}
\altaffiltext{1}{Department of Astronomy, Kyoto University, Kitashirakawa-Oiwake-cho, Sakyo, Kyoto 606-8502, Japan}
\altaffiltext{2}{Okayama Branch Office, Subaru Telescope, National Astronomical Observatory of Japan, NINS, 3037-5 Honjo, Kamogata, Asakuchi, Okayama 719-0232, Japan}
\altaffiltext{3}{Astronomical Observatory, Kyoto University, Kitashirakawa-Oiwake-cho, Sakyo, Kyoto 606-8502, Japan}
\altaffiltext{4}{Department of Social Data Science, Hitotsubashi University, 2-1 Naka, Kunitachi, Tokyo 186-8601, Japan}
\altaffiltext{5}{Department of Multi-Disciplinary Sciences, Graduate School of Arts and Sciences, The University of Tokyo, 3-8-1 Komaba, Meguro, Tokyo 153-8902, Japan}
\altaffiltext{6}{Kwasan Observatory, Kyoto University, 17 Ohmine-cho, Kita-Kazan, Yamashina, Kyoto 607-8471, Japan}
\altaffiltext{7}{Department of Environmental Systems Science, Doshisha University, 1-3 Tataramiyakodani, Kyotanabe, Kyoto 610-0394, Japan}
%\footnotetext[$\dag$]{Present address: ....}

%%% end:list of authors

%% !!! Select 3 to 5 words from PASJ's key words !!! 
%% List of Key Words: https://academic.oup.com/pasj/pages/Pasj_Keywords 
%% "\KeyWords{ }" always has to be placed before ``\maketitle'' 

\maketitle

\begin{abstract}
  We have developed and commissioned MIDSSAR (MId Dispersion Spectrograph for Stellar Activity Research), a novel fiber-fed, dual-band spectrograph optimized for high-cadence monitoring of stellar superflares and magnetic activity phenomena, mounted on the 3.8-m Seimei Telescope at Kyoto University Okayama Observatory. Utilizing a custom-designed volume phase holographic grating and a commercially available volume binary transmission grating, the instrument achieves simultaneous observations of key chromospheric activity indicators: the Ca~{\sc ii} H\&K band ($382.5\mbox{--}406.5$~nm) at a resolving power of $R = 13,800$, and the H$\alpha$ band ($641.1\mbox{--}667.6$~nm) at $R = 18,200$. MIDSSAR is strategically integrated with the existing high-dispersion spectrograph GAOES-RV to enable concurrent photosphere-chromosphere monitoring. The control software coordinates high-speed dual-camera exposures down to $1\text{ ms}$ while incorporating real-time telescope telemetry into the data headers. We also developed a dedicated data reduction pipeline that automatically executes calibration, spatial reconstruction of the 61-fiber bundle, and one-dimensional spectral extraction with a wavelength precision of $\sim 0.002\text{ nm}$. Initial on-sky test observations conducted from March to May 2026 successfully verified the system's quantitative readiness. The observed signal-to-noise ratio characteristics perfectly align with modeled theoretical limits under varying seeing conditions, confirming that the entire instrument complex successfully meets its primary design requirements and operational expectations for systematic stellar astrophysics surveys.

\end{abstract}

%\pagewiselinenumbers 

\section{Introduction}
Solar and stellar flares are intense explosions in stellar atmospheres driven by the release of magnetic energy around sunspots and starspots, and are sometimes accompanied by plasma eruptions (for reviews, see \cite{Shibata2011}; \cite{Kowalski2024}).
The properties of stellar flares and associated plasma eruptions have primarily been investigated using the temporal variations of H$\alpha$ lines (e.g., \cite{Honda2018}; \cite{Namekata2020}; \cite{Maehara2021}; \cite{Inoue2023}; \cite{Namizaki2023}; \cite{Notsu2024}; \cite{Ichihara2025}; \cite{Kajikiya2025}) in comparison with the solar ones (e.g., \cite{Otsu2022}, \yearcite{Otsu2024}, \yearcite{Otsu2026}).
While many superflares have been reported on solar-type stars (e.g., \cite{Maehara2012}; \cite{Shibayama2013}; \cite{Maehara2015}; \cite{Notsu2019}; \cite{Okamoto2021}), filament/prominence eruptions associated with superflares have recently been discovered on solar-type stars (\cite{Namekata2021}, \yearcite{Namekata2024a}, \yearcite{Namekata2025}, \yearcite{Namekata2026}) primarily using low-dispersion spectroscopy with KOOLS-IFU (\cite{Matsubayashi2025}) on the 3.8-m Seimei Telescope at Kyoto University (\cite{Kurita2020}).
However, the modest spectral resolution of KOOLS-IFU ($R \sim 2000$) can mask slow plasma motions, limiting our ability to detect detailed kinematic signatures; thus, higher spectral resolution is urgently required.

Superflares are driven by large starspots and magnetic fields on the stellar surface (e.g., \cite{Notsu2013b}; \cite{Maehara2017}; \cite{Ikuta2023}), making it crucial to investigate the direct relationship between superflare activity and starspot properties as observed in both photospheric and chromospheric lines (\cite{Notsu2013a}; \cite{Nogami2014}; \cite{Notsu2015a}; \yearcite{Notsu2015b}; \cite{Honda2015}; \cite{Ikuta2026}).
The properties of such starspots have already been successfully explored via Doppler imaging using GAOES-RV (\cite{Sato2024}) on the Seimei Telescope (\cite{Lee2026}).
However, because GAOES-RV is a high-dispersion spectrograph ($R \sim 65,000$) designed specifically for Doppler planet searches with a wavelength coverage restricted to $516\mbox{--}593\text{ nm}$, it cannot capture the primary chromospheric diagnostic lines---namely H$\alpha$ and Ca$\,\text{II}$ H\&K---despite its capability to observe certain chromospheric absorption lines with long exposure times.
Therefore, to investigate superflares and plasma eruptions with superior velocity resolution, and to examine the direct connection between superflares and the chromospheric lines originating from active starspots, we have developed \textbf{MIDSSAR} (\textbf{MI}d \textbf{D}ispersion \textbf{S}pectrograph for \textbf{S}tellar \textbf{A}ctivity \textbf{R}esearch).

In addition to its primary capability of detecting eruptive phenomena, MIDSSAR is expected to provide an unprecedented multi-layer diagnostic of the lower stellar atmosphere during energetic flare events.
Specifically, mid-dispersion coverage of H$\alpha$ and Ca$\,\text{II}$ H\&K lines is useful to capture simultaneous line-profile variations, such as redshifted asymmetries and Doppler broadenings, which serve as direct empirical signatures of chromospheric condensation downflows (e.g., \cite{Kowalski2015}).
Such comprehensive observational constraints are particularly essential for rigorously testing and calibrating state-of-the-art radiative hydrodynamic (RHD) models (e.g., \cite{Allred2015}; \cite{Kowalski2017}), thereby advancing our fundamental understanding of atmospheric energy transport mechanisms. Furthermore, by precisely characterizing the dynamic radiation environment and high-energy emission profiles during flares, MIDSSAR enables a robust evaluation of stellar space weather impacts, including upper-atmospheric heating, photo-evaporation, and driven atmospheric escape on orbiting exoplanets (e.g., \cite{Lammer2003}; \cite{Ribas2005}; \cite{Airapetian2020}).
Consequently, MIDSSAR is expected to serve as a highly versatile and integrative observational platform that directly bridges the gap between fundamental stellar magnetohydrodynamics and exoplanetary science.

At the focus of the Seimei Telescope, incoming stellar light is split by wavelength: the mid-visible band ($500\mbox{--}600\text{ nm}$) is fed into GAOES-RV for precise photospheric line-profile analysis, while the remaining ultraviolet/blue and red wavelengths are routed to MIDSSAR to capture the Ca$\,\text{II}$ H\&K lines (393.4, 396.9\>nm) and the H$\alpha$ line (656.3\>nm) simultaneously.
Operating as a single, integrated spectrograph complex, this setup allows us to simultaneously monitor the photosphere and chromosphere across multi-resolution bands.

The primary targets for the MIDSSAR+GAOES-RV complex are RS CVn-type binaries and rapidly rotating young solar-type stars that exhibit vigorous magnetic activity and superflares, alongside quiet solar-type stars required for baseline comparisons. In the northern sky, there are approximately 30, 70, and 130 RS CVn systems with magnitudes brighter than $V < 7$, $8$, and $9\text{ mag}$, respectively, with a comparable number of young solar-type stars and several tens of times more quiet solar-type stars. Among these, prominent flaring targets previously investigated with KOOLS-IFU include the RS CVn stars V1355~Ori ($V = 9.2\text{ mag}$; \cite{Inoue2023}) and UX~Ari ($V = 6.4\text{ mag}$; \cite{Urabe2026}), as well as the young solar-type stars EK~Dra ($V = 7.6\text{ mag}$) and V889~Her ($V = 7.4\text{ mag}$) (\cite{Namekata2025}); these systems represent key objects for ongoing MIDSSAR monitoring. To resolve the typical durations of stellar superflares (ranging from tens of minutes to a few hours), the instrument requirement dictates achieving $S/N > 100$ for a $V \sim 9\text{ mag}$ star within a few-minute exposure time (cf. \cite{Notsu2024}; \cite{Namekata2025}). Furthermore, detecting plasma kinematic signatures of $\sim 20\text{ km s}^{-1}$ associated with solar and stellar flares---such as chromospheric condensation downflows and post-flare loops---requires simultaneous observation of both the $\text{H}\alpha$ and $\text{Ca}\,\text{II}$ H\&K lines at a spectral resolution of $R \sim 15,000$ (cf. \cite{Kowalski2015}; \cite{Notsu2024}; \cite{Otsu2024}). In addition, real-time equivalent width (EW) measurement capabilities are required to identify flare onsets during observations and secure uninterrupted, high-cadence time-series datasets throughout the flaring event.

Globally, six high dispersion spectrographs capable of observing the $\text{Ca}\,\text{II}$ H\&K lines are operational on telescopes larger than the 3.8-m Seimei Telescope. However, all of these are general-purpose high-dispersion spectrographs covering the entire optical band. Due to the inherent efficiency loss of $p$-polarized light on reflective echelle gratings, as well as operating far from the peak efficiency wavelength of cross-dispersers, the inherent advantages of larger collecting areas and superior site transparency are significantly degraded at these blue wavelengths. Moreover, when monitoring short-timescale variations on the order of minutes, large-format CCD detectors impose substantial readout overheads ($\sim 30\text{ s}$ per frame), which further impairs their operational efficiency for rapid time-series spectroscopy. Furthermore, high user demand on large-aperture facilities severely restricts long-term, dedicated spectroscopic monitoring needed to catch infrequent events such as superflares on young solar-type stars, which are traditionally conducted on telescopes of $\le 2\text{-m}$ aperture. While mid-dispersion ($R \sim 10,000$) spectrographs on $2\text{-m}$ class telescopes are used worldwide to monitor flares on M-dwarfs, achieving the high $S/N$ required for solar-type superflares remains prohibitive with such smaller facilities. MIDSSAR overcomes these limitations by employing high-efficiency transmission gratings (even for $p$-polarization) and fast CMOS detectors with near-zero readout time. Combined with the 3.8-m Seimei Telescope and operated simultaneously with GAOES-RV, MIDSSAR establishes an internationally unique and optimal platform optimized for long-term spectroscopic monitoring of stellar flares.

This paper is organized as follows. In Section 2, we present the optical, mechanical, and software design of MIDSSAR, highlighting its seamless integration and co-observation scheme with the high-dispersion spectrograph GAOES-RV at the 3.8-m Seimei Telescope.
Section 3 details the data reduction pipeline, calibration processes, and the subsequent quantitative performance verifications evaluated through our initial engineering and test observations.
Finally, a summary and the future prospects of this instrument are provided in Section 4.

\section{Optical Design}\label{sec:2}

\subsection{Spectrograph Main Body}\label{ssec:21}

The main body of MIDSSAR is designed as a high-efficiency, dual-band spectrograph that processes the ultraviolet/blue and red channels simultaneously using a combination of a custom-designed volume phase holographic (VPH) grating and a commercially available volume binary (VB) transmission grating.

The optical layout begins with a fiber slit composed of 61 fibers, each featuring a $50\,\mu\text{m}$ core diameter, arranged linearly. The beam emerges from this fiber slit with a focal ratio of $F/4.2$ and is collimated by an on-axis collimator with a focal length of $560\text{ mm}$. The collimated beam is then reflected by a perforated plane mirror to direct the light toward the subsequent splitting and dispersing optics.

Downstream of the perforated mirror, a dichroic mirror splits the incoming beam into the two primary operational channels:
\begin{itemize}
    \item \textbf{Ultraviolet/Blue Channel (Ca~\textsc{ii} H\&K):} The light in the shorter wavelength regime is reflected by the dichroic mirror to cover the wavelength range of $382.5\mbox{--}406.5\text{ nm}$. It passes through a $\phi135\text{ mm}$ VPH grating manufactured by Wasatch Photonics (custom-fabricated with a groove density of $2532\text{ lines mm}^{-1}$ optimized for an angle of incidence of $30^\circ$ at a central wavelength of $395\text{ nm}$). The dispersed light is then focused by a dedicated blue camera lens system consisting of three elements. This system is divided into two groups: a closely spaced doublet (convex and concave lenses) for primary focusing, and a single correction lens positioned well downstream, immediately in front of the detector. The first element is made of Ohara S-FSL5 glass, while the remaining two elements utilize Ohara S-NBH52V glass. 
    \item \textbf{Red Channel (H$\alpha$):} The longer wavelength light transmits through the dichroic mirror to cover $641.1\mbox{--}667.6\text{ nm}$. It enters a $\phi135\text{ mm}$ Volume Binary (VB) transmission grating manufactured by LightSmyth with a groove density of $1850\text{ lines mm}^{-1}$. Although this commercial grating is nominally designed for $800\text{ nm}$ at an angle of incidence (AOI) of $47.8^\circ$, we operate it at an AOI of $37.4^\circ$ for $656\text{ nm}$, which achieves a high theoretical diffraction efficiency of $\sim 90\%$ (confirmed by the manufacturer). The beam is focused by a dedicated red camera lens system. While it shares the identical glass sequence as the blue channel (an Ohara S-FSL5 element followed by two Ohara S-NBH52V elements) and utilizes the same split-group configuration, its detailed geometric optical parameters are independently optimized to achieve the best imaging performance at the red wavelengths.
\end{itemize}

Given that the optical design is relatively straightforward and standard for this class of instruments, the prescription of the glass materials alone sufficiently defines the system's characteristics without requiring the detailed enumeration of the individual optimization parameters.

Both channels utilize identical camera systems: the BH-67M manufactured by BITRAN, equipped with a Canon LI3030SAM back-illuminated, large-area CMOS sensor. This detector features a pixel pitch of $19\,\mu\text{m} \times 19\,\mu\text{m}$, a total pixel resolution of $2160 \times 1280$, and a maximum frame transfer rate of $98\text{ fps}$. On each detector, the 61 distinct fiber spectra are captured simultaneously within a central $300\text{-pixel}$ wide region of interest (ROI). This high-quantum-efficiency, large-format detector configuration ensures optimal spatial and spectral sampling, satisfying the target resolving powers of $13,800$ for the Ca~\textsc{ii} H\&K band and $18,200$ for the H$\alpha$ band.

\begin{figure*}
 \begin{center}
  \includegraphics[width=13cm]{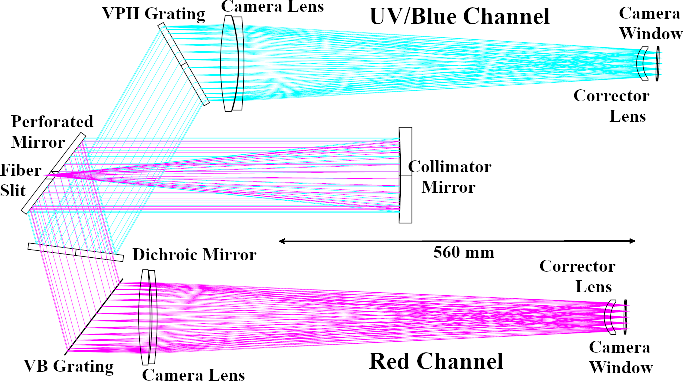} 
 \end{center}
\caption{ Optical layout of the spectrograph main body. Light entering from the fiber slit at the left end is collimated and then divided into two wavelength bands, which are subsequently dispersed by two distinct gratings and focused by their respective camera lens systems.
 {Alt text: An optical ray-tracing diagram showing a parabolic mirror, a perforated plane mirror, a dichroic mirror, two diffraction gratings, and two sets of camera lens systems, illustrating the specific light paths originating from the fiber slit. } 
}\label{fig:optical_layout}
\end{figure*}

\begin{figure}
 \begin{center}
  \includegraphics[width=8cm]{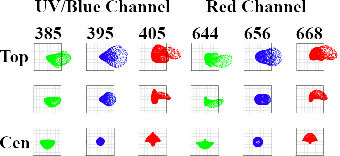} 
 \end{center}
\caption{ Spot diagrams of the spectrograph. For the two channels, the spots represent light incident from the center, the top edge, and the intermediate position of the fiber slit, evaluated at the center and both ends of each wavelength band (wavelengths are labeled within the panels). Spots for light from the bottom half of the slit are symmetric to those from the top half. The squares indicate the size of a single pixel ($19\,\mu\text{m}$).
 {Alt text: A spot diagram showing the geometric image shapes formed by light entering from three distinct slit positions, plotted for six different wavelengths. } 
}\label{fig:spot_diagram}
\end{figure}

Figure~\ref{fig:optical_layout} illustrates the complete optical layout of the spectrograph main body, and Figure~\ref{fig:spot_diagram} presents the resulting spot diagrams across the entire focal plane for both channels. The optical system delivers exceptional imaging performance that satisfies the Nyquist-Shannon sampling criterion. Specifically, the projected image of the $50\,\mu\text{m}$ fiber core on the CMOS detector corresponds to slightly less than $3\text{ pixels}$ ($\approx 2.6$ and $2.9\text{ pixels}$ in the UV/Blue and the Red channels, respectively), providing an optimal balance between spectral resolution and sampling efficiency. 

From a design philosophy perspective, the $50\,\mu\text{m}$ physical diameter of the fiber core is approximately 20 times larger than the diffraction-limited spot size ($\sim 2.5\,\mu\text{m}$) at a focal ratio of $F/4.2$. Consequently, the strict requirement for diffraction-limited imaging performance is significantly relaxed, meaning that an imaging precision within a few times the diffraction limit is more than sufficient for this instrument. Indeed, the geometric optical aberrations are strictly suppressed to within $1\text{ pixel}$ ($19\,\mu\text{m}$) at all positions across the active area of the detectors. This negligible aberration level ensures that the instrumental profile is overwhelmingly dominated by the physical fiber slit width rather than optical imperfections, guaranteeing stable and symmetrical line profiles crucial for precise line-asymmetry measurements during superflare events.

\subsection{Fore-optics}\label{ssec:22}

The fore-optics system of MIDSSAR is placed at the telescope focus and is strictly designed to interface seamlessly with the existing high-dispersion spectrograph GAOES-RV, while overcoming severe atmospheric dispersion at low elevation angles.

\subsubsection{Design Constraints and Operation Scheme}
When GAOES-RV is operational, the telescope's instrument rotator remains fixed. This operational approach is primarily mandated because the fore-optics system of GAOES-RV is too large to be mounted directly onto the moving rotator assembly, and maintaining a fixed configuration also ensures mechanical stability. Furthermore, since the primary science targets are point sources, the resulting field rotation during an observation does not compromise the spectroscopic data quality.

However, locking the rotator directly affects the behavior of atmospheric dispersion at the telescope focus. Because the rotator does not track the sky, the field of view undergoes continuous rotation relative to the horizon. Consequently, the dynamic rotation of the atmospheric dispersion vector at the telescope focus is determined by the combination of this field rotation and the changing relative orientation between the celestial north and the alt-azimuth directions. Geometrically, these two rotational components effectively combine such that the orientation of the atmospheric dispersion vector in the field of view simply tracks the telescope's zenith distance (or altitude angle). At a low elevation angle of $20^\circ$, this combined geometric and atmospheric dispersion causes the image position at the UV/Blue channel wavelengths to shift by more than $3''$ away from the $550\text{ nm}$ baseline.

Simultaneously, precise auto-guiding must be maintained using the GAOES-RV guiding system. Since GAOES-RV is also a fiber-fed spectrograph, its guiding camera functions as a hole-viewer rather than a traditional slit-viewer, monitoring a small circular aperture that actively feeds the instrument's science fiber. To ensure stable throughput, the target star's position must be locked at the central guiding wavelength of $550\text{ nm}$ exactly within this circular aperture. Because the atmospheric dispersion vector continuously drifts and expands beyond the $\phi5''$ field of view of the MIDSSAR fiber bundle, an atmospheric dispersion corrector (ADC) is absolutely indispensable to prevent severe chromatic photon loss.

Through rigorous optical optimization, we discovered that a pair of identical Amici (direct-vision) prisms---each consisting of a $\text{CaF}_2$ and an Ohara S-LAH97 element optimized at $550\text{ nm}$---can successfully compensate for this time-varying dispersion, dynamically correcting the positional displacements of both the UV/Blue and Red channels relative to the $550\text{ nm}$ guiding center when counter-rotated.

\subsubsection{Optical Components and Layout}
The existing collimator lens for GAOES-RV exhibits significant chromatic aberration, making it unsuitable for the wide-wavelength coverage required by MIDSSAR. To resolve this, a new achromatic triplet collimator lens was designed. This triplet comprises an Ohara S-FSL5 plano-convex lens glued to an Ohara S-NBH52V concave lens, paired with a $\text{CaF}_2$ plano-convex lens placed across a small air gap. During the optimization process, the outermost incident and exit surfaces converged toward almost perfectly flat profiles. Consequently, we locked both external surfaces to be perfectly flat, executing a final optimization that yields a highly robust, easy-to-manufacture solution. 

\begin{figure*}
 \begin{center}
  \includegraphics[width=14.5cm]{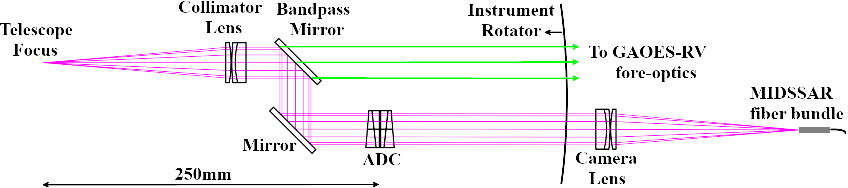} 
 \end{center}
 \caption{ Optical element layout of the MIDSSAR fore-optics system at the telescope focus. The F/6 incident beam from the telescope passes through the collimator lens and is divided by the bandpass mirror into the beam directed toward the GAOES-RV fore-optics system and the remaining wavelengths. The latter passes through the ADC and is subsequently focused by the camera lens, which is mounted on the fixed structure outside the instrument rotator, to form an image onto the fiber bundle face.
 {Alt text: An optical ray-tracing diagram showing the layout of the collimator lens, the bandpass mirror, the counter-rotating ADC prisms, and the camera lens on the fixed structure, illustrating how the light path splits and focuses onto the fiber bundle.}
}\label{fig:fore_optics_layout}
\end{figure*}

\begin{figure}
 \begin{center}
  \includegraphics[width=8cm]{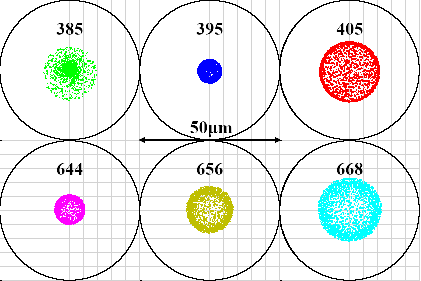} 
 \end{center}
\caption{ Spot diagrams evaluated after the fore-optics correction. The circle indicates the physical core diameter ($50\,\mu\text{m}$) of the fiber, and the numbers labeled on each spot denote the corresponding wavelengths. Depending on the relative counter-rotation angles of the ADC prisms, the centers of these spots shift from being perfectly aligned to being appropriately separated in the direction that compensates for atmospheric dispersion.
 {Alt text: A spot diagram showing the geometric image shapes at six different wavelengths for the incident light at the center of the field of view.}
}\label{fig:fore_optics_spots}
\end{figure}

The complete optical path of the fore-optics proceeds as follows (Figure~\ref{fig:fore_optics_layout}):
\begin{enumerate}
    \item The divergent beam from the telescope is collimated by the newly designed achromatic triplet lens.
    \item A bandpass mirror placed downstream reflects the UV/Blue and Red wavelengths toward MIDSSAR, while transmitting only the $500\mbox{--}600\text{ nm}$ band to feed the GAOES-RV fore-optics system.
    \item The reflected beam passes through the counter-rotating ADC prisms to compensate for the atmospheric dispersion.
    \item Finally, an imaging triplet lens---identical in design to the triplet collimator lens described above---re-focuses the corrected beam onto the input face of the MIDSSAR fiber bundle.
\end{enumerate}

Figure~\ref{fig:fore_optics_layout} displays the optical element layout, and Figure~\ref{fig:fore_optics_spots} shows the resulting spot diagrams on the fiber face. 

It has been verified that replacing the original GAOES-RV collimator with this new triplet lens introduces no degradation to the GAOES-RV spectroscopic capabilities. While the splitting of the beam significantly narrows the wavelength coverage available to the guiding camera compared to the previous full-wavelength operation, the resulting photon loss does not critically impair the guiding performance. This resilience is due to the fact that the guider CCD's quantum efficiency peaks near $550\text{ nm}$, ensuring that a sufficient signal-to-noise ratio is maintained for stable auto-guiding within the circular fiber aperture during co-observation.

\subsection{Fiber Bundle}\label{ssec:23}
For high-resolution spectrographs targeting stellar point sources, a single large-core multimode fiber is commonly employed. However, MIDSSAR adopts a bundle of small-core multimode fibers due to several key mechanical and operational motivations:
\begin{itemize}
    \item \textbf{Simplification of the Telescope Interface:} A single-fiber optical feed generally requires a dedicated focal-plane viewer system for target acquisition and focus adjustment. In our setup, auto-guiding is performed by the pre-existing viewer camera of GAOES-RV. Consequently, the input unit of MIDSSAR only requires a simple mechanism for three-dimensional positioning---namely, aligning the target image laterally and adjusting its focus relative to the fiber input end. Adopting a fiber bundle bypasses the need for an additional viewer camera at the input feed, significantly simplifying the mechanical structure. This compact design was crucial because MIDSSAR was retrofitted into the existing focal mechanism of GAOES-RV.
    \item \textbf{Simplification of the Spectrograph Input Optics:} To achieve the targeted spectral resolution with a single large-core fiber, an $F$-ratio conversion system and an image slicer would be required at the spectrograph entrance. Using a bundle of small-core fibers functionally acts as a spatial slicer, eliminating the need for additional slicing optics.
    \item \textbf{Background Light Subtraction Capability:} Although background contamination is generally negligible during routine observations, the multi-fiber configuration provides operational flexibility in exceptional cases---such as observations under strong moonlight or during twilight/dawn. Under these special conditions, peripheral fibers surrounding the central stellar image can be used to sample and subtract the sky background.
\end{itemize}

The fiber bundle consists of 61 individual high-grade optical fibers (Optran UV, manufactured by CeramOptec), featuring a numerical aperture (NA) of 0.12, a physical core diameter of $50\,\mu\text{m}$, and a total length of $6\text{ m}$. At the telescope focus, these 61 science fibers are arranged in a closely packed hexagonal configuration to maximize the spatial filling factor. Conversely, at the spectrograph input, they are linearly aligned to form a single fiber slit with a constant fiber-to-fiber spacing of $61\,\mu\text{m}$. To facilitate precise wavelength calibration, an additional reference fiber with a length of $1\text{ m}$ is positioned immediately adjacent to each longitudinal end of this fiber slit. These two calibration channels are illuminated by emission-line light sources at appropriate intervals during observations.

Due to the fundamental mechanical difficulties associated with precisely manipulating and arranging fibers with a fine $50\,\mu\text{m}$ core diameter, achieving a perfectly idealized close-packed configuration at the telescope-side bundle face proved exceptionally challenging. As shown in Figure 5, the fabricated bundle face exhibits significant geometric irregularities, most notably on the right and bottom sides where the fibers appear disordered and piled up, disrupting the overall hexagonal structure. Because the science targets are expected to be positioned near the relatively dense central region of the bundle during observations, the effective filling factor for the scientific data is approximately 60\%, whereas the filling factor in the highly distorted peripheral regions is lower than this value. Despite this pronounced structural imperfection, the spatial arrangement remains highly stable, ensuring robust and consistent wavelength transmission characteristics over time.

\begin{figure}
 \begin{center}
  \includegraphics[width=8cm]{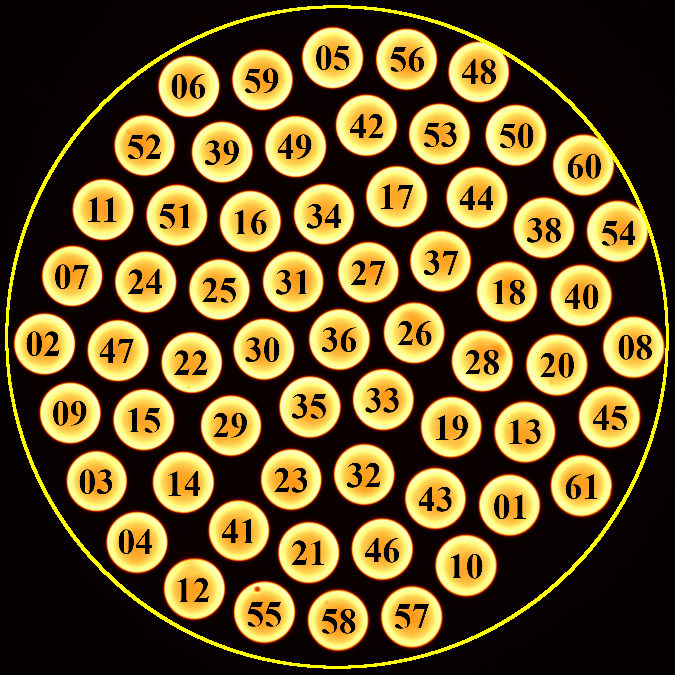} 
 \end{center}
\caption{ Telescope-side end face of the fiber bundle. The numbers in the diagram denote the sequential index starting from the edge of the linear fiber slit on the spectrograph side, demonstrating that the fibers near the center of the slit are arranged to cluster around the center of the bundle face. The outer enclosing circle represents a physical diameter of $550\,\mu\text{m}$, which corresponds to a $5''$ field of view on the sky.
 {Alt text: An enlarged image of the fiber bundle end face showing the arrangement of 61 individual fibers.}
}\label{fig:fiber_layout}
\end{figure}

The specific spatial mapping between the linear fiber slit and the hexagonal bundle face is indexed in Figure~\ref{fig:fiber_layout}. It is noteworthy that the fibers positioned near the center of the fiber slit (around fiber number 30) are successfully localized within the central region of the telescope-side bundle face. This strict spatial correspondence is critical for the data reduction process; it allows the user to reconstruct the two-dimensional spatial image profile on the sky by synthesizing the individual, resolved spectra of the 61 fibers captured on the spectrograph detector.

\subsection{Mechanical Structure and Thermal Control}\label{ssec:24}

\begin{figure*}
 \begin{center}
  \includegraphics[width=13cm]{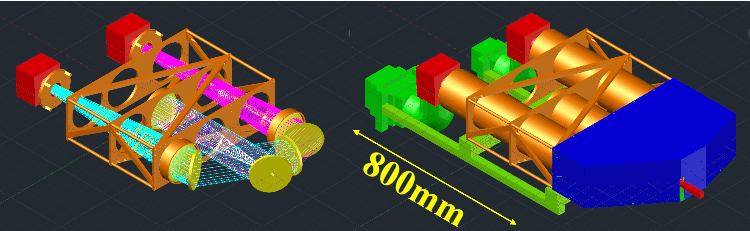} 
 \end{center}
 \caption{ Mechanical structure of the spectrograph. (Left) Internal layout showing the optical elements and light path. (Right) Entire assembled mechanical structure. The relative focal positions of the two cameras are fixed, and the global focusing of the system is executed by translationally moving the fiber slit along the optical axis.
 {Alt text: Two 3D CAD drawings of the spectrograph: the left panel shows the internal optical layout and light path, and the right panel shows the fully assembled system with its external covers.}
}\label{fig:spectrograph_structure}
\end{figure*}

Although MIDSSAR is a fiber-fed spectrograph, the total fiber length is strictly constrained to $6\text{ m}$ to minimize photon loss at ultraviolet (UV) wavelengths. Consequently, the entire spectrograph body is designed to be suspended directly beneath the Nasmyth platform of the telescope. This specific location subjects the instrument to significant ambient temperature fluctuations during observations, which can cause thermal focal shifts within the spectrograph and drifts in the spectral positions on the detector. To mitigate these thermal effects, the main mechanical components are fabricated from stainless steel. In particular, the tube housing the camera lens assembly is constructed from ferritic stainless steel, which features a low coefficient of thermal expansion (CTE). As illustrated in Figure~\ref{fig:spectrograph_structure}, the structural design achieves overall mass reduction by employing a box-like structure for the section spanning from the fiber slit to the diffraction grating, while adopting cylindrical structures for the camera lens and collimator mirror sections. The instrument maintains a highly simplified architecture with only two moving mechanisms: the input slit assembly, which moves translationally along the optical axis for focusing, and a shutter located immediately behind it.

The thermal management system is also optimized to ensure stable detector performance. If the ambient temperature inside the camera housing exceeds $40^\circ\text{C}$, achieving the required detector cooling temperature of $-20^\circ\text{C}$ becomes challenging. To address this, an external Peltier cooling unit is integrated into the camera system to actively regulate the internal temperature when necessary. For stray light mitigation, the interior surfaces of the spectrograph housing are coated with NEXTEL Velvet-Coating, which exhibits an exceptionally low reflectance of 0.02\% in the visible band. Furthermore, to prevent condensation on the camera input window, the instrument incorporates a dry-air purge system. Dry air is continually introduced from the side of the final lens holder toward the camera input window, successfully filling the entire interior of the lens tube with a dry environment.

\subsection{Fore-optics Mechanical Structure}\label{ssec:25}

\begin{figure}
 \begin{center}
  \includegraphics[width=8cm]{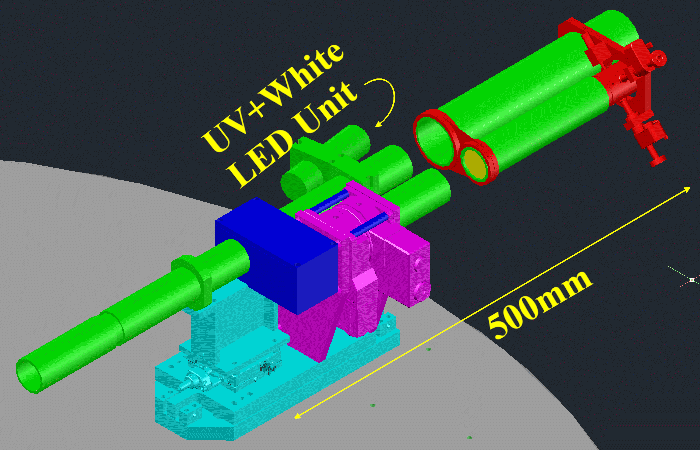} 
 \end{center}
\caption{ Mechanical structure of the fore-optics. It is fabricated from an aluminum alloy and incorporates a commercially available $\phi 30\text{ mm}$ lens tube and two automated rotation stages for the ADC. Rotating the instrument rotator by $8^\circ$ positions the LED flat-field calibration light source unit directly opposite the imaging lens.
 {Alt text: A 3D CAD drawing showing two sets of parallel-arranged tubes integrated with two automated rotation stages on a manual translational stage.}
}\label{fig:fore_optics_structure}
\end{figure}

The fore-optics system is built as a compact and rigid assembly to ensure precise beam delivery to the fiber bundle. As shown in Figure~\ref{fig:fore_optics_structure}, it comprises a dedicated lens tube with an inner diameter of $30\text{ mm}$, a bandpass mirror housing box, two motorized counter-rotating ADC stages, and a two-axis Picomotor actuator system dedicated to the fine positional alignment of the fiber bundle. The entire assembly is mounted atop a multi-axis positioning stage that aligns the instrument input with the telescope focal plane. The section from the camera lens onward is designed to be securely suspended directly beneath the baffle tube of the GAOES-RV fore-optics system. Furthermore, the instrument rotator can be intentionally offset by $8^\circ$, which brings a dedicated calibration lamp unit into position directly in front of the camera lens. Because the light emitted from this UV+white LED unit is diffused through a very thin diffuser sheet, the calibration system does not require strict optical alignment; simply positioning the lamp unit roughly in front of the camera lens is sufficient to fill the input beam path. This loose tolerance is further aided by the fact that the input face of the fiber bundle is exceptionally small (less than $0.5\text{ mm}$ in diameter). Rather than employing a thick, high-extinction diffusing plate, the choice of a thin sheet successfully minimizes photon loss, maximizing the throughput of the calibration light. This highly efficient design allows for high-cadence, short-exposure sequences, enabling the rapid acquisition of a large number of frames to achieve a high signal-to-noise ratio ($S/N$) for flat-field calibrations during operational gaps.

\subsection{Control Software}\label{ssec:26}

\begin{figure}
 \begin{center}
  \includegraphics[width=8cm]{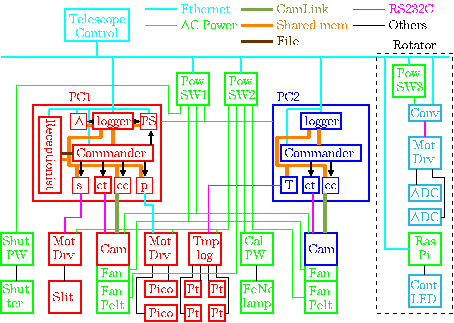}
 \end{center}
 \caption{System block diagram of the MIDSSAR control software.
 In each PC, processes enclosed by thick-lined boxes represent resident (daemon) processes, whereas those enclosed by thin-lined boxes represent processes invoked asynchronously. The abbreviations used in the diagram are defined as follows:
(A)~ADC control,
(PS)~power supply control,
(s)~slit position control,
(ct)~detector temperature control,
(cc)~camera exposure control,
(p)~fore-optics bundle positioning control,
(T)~instrument temperature logger,
(Pow~SW)~network power switch,
(Pelt)~Peltier cooler,
(Pico)~Picomotor actuator,
(Pt)~platinum resistance thermometer,
(FeNe~lamp)~Fe-Ne hollow cathode lamp,
(Conv)~Ethernet-to-RS232C converter,
(Ras~Pi)~Raspberry Pi computer for auxiliary processing on the rotator, and
(Cont~LED)~UV+white LED source for flat-field calibration.
 {Alt text: A comprehensive system schematic showing all hardware and software components connected by lines that indicate their respective communication protocols.}
}\label{fig:control_software}
\end{figure}

The minimum exposure time for the two CMOS cameras of MIDSSAR is $1\text{ ms}$, and to achieve stable image acquisition during high-speed continuous exposures, each camera must be driven by its own dedicated computer. On the other hand, GAOES-RV, which is operated concurrently with MIDSSAR, utilizes a CCD detector that is physically incapable of high-speed continuous frame rates. Consequently, during standard observations with exposure times tuned to GAOES-RV, timing precision between the instruments is practically negligible. However, for specialized observations such as monitoring stellar flares, MIDSSAR may be operated in a high-speed continuous exposure mode, requiring the two CMOS cameras to be driven with a certain degree of temporal synchronization. Under such high-speed operations, collecting all peripheral statuses---primarily consisting of telescope telemetry and environmental logs such as temperature, humidity, and atmospheric pressure---at the start and end of every single exposure would impose a significant load on the network, potentially compromising operational stability. To ensure stable performance even during high-speed exposures, the control system of MIDSSAR (Figure~\ref{fig:control_software}) has been developed based on the following specific policies.

\begin{itemize}
    \item Each PC runs two distinct core background tasks: a ``logger'' process that samples status information periodically at a 1-Hz frequency, and a ``commander'' process that executes operational commands asynchronously.
    \item The vast majority of the global telemetry collection is performed by Logger~1 (on PC~1) at 1-Hz intervals. The gathered data are then transmitted directly to Logger~2 (on PC~2). Conversely, Logger~2 monitors the localized process states within its own host PC and communicates this internal status back to Logger~1.
    \item PC~1 runs a dedicated ``receptionist'' process that constantly listens for incoming commands. Upon receiving a new command sequence, the receptionist assigns a unique command ID to each instruction and generates an individual file for each command on the local disk.
    \item As soon as Commander~1 detects a new command file, it executes the tasks sequentially in order of their command IDs. If a task requires a process managed by PC~2, Commander~1 automatically forwards the specific command to Commander~2.
    \item Within each individual PC, a shared memory segment is implemented, allowing all local processes to instantaneously share and access the entire system status and telemetry data without inter-process communication overhead.
\end{itemize}
Under these schemes, the two cameras are controlled with a loose hardware/software synchronization of approximately $100\text{ ms}$. This level of synchronization is more than sufficient for standard flare monitoring observations, where individual exposure times typically range from tens of seconds to several minutes, as well as for high-cadence observations down to a few seconds. Even when conducting extremely rapid time-series observations with exposure times down to a few milliseconds, precise time stamps calibrated via the Network Time Protocol (NTP) with sub-millisecond accuracy ($< 1\text{ ms}$) are embedded into each data header, allowing the data pipeline to successfully match and pair corresponding frames from simultaneous exposures during post-processing. Using this highly optimized and robust software architecture, an observer can advance the entire observation run smoothly. By simply sending a pre-defined command sequence (such as a focusing routine) to the receptionist process, all sub-systems automatically coordinate, execute their respective tasks in sequence, and securely save the scientific data.

\section{Data Reduction and Performance Verification}\label{sec:3}

\begin{figure}
 \begin{center}
  \includegraphics[width=8cm]{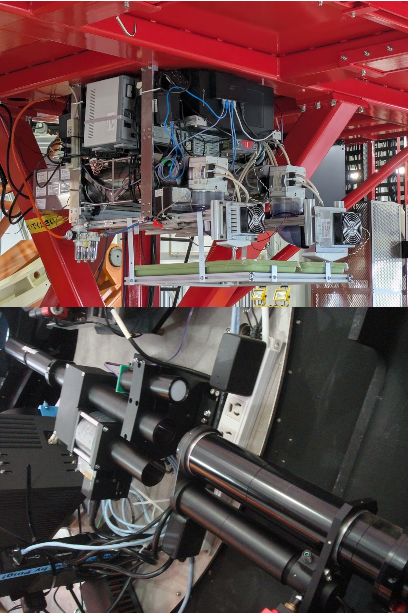}
 \end{center}
\caption{ Photographs of MIDSSAR mounted on the Seimei Telescope. (Top) The main body of the spectrograph installed beneath the Nasmyth platform. (Bottom) The MIDSSAR fore-optics unit mounted on the instrument rotator and positioned below the GAOES-RV fore-optics baffle.
 {Alt text: Two photographs showing the MIDSSAR instrument components as installed on the telescope, with the upper panel displaying the spectrograph main body and the lower panel showing the fore-optics system.}
}\label{fig:seimei_mount}
\end{figure}

MIDSSAR was successfully mounted on the Seimei Telescope in February 2026 (Figure~\ref{fig:seimei_mount}). This chapter detailedly describes the various calibration frames required for accurate image analysis, the comprehensive architecture of the data reduction pipeline, and the subsequent quantitative performance verification evaluated through initial test observations.

\subsection{Instrument Alignment and Calibration Frames}\label{sec:31}

\subsubsection{Focus Alignment and Anchor Images}\label{ssec:311}

\begin{figure}
 \begin{center}
  \includegraphics[width=8cm]{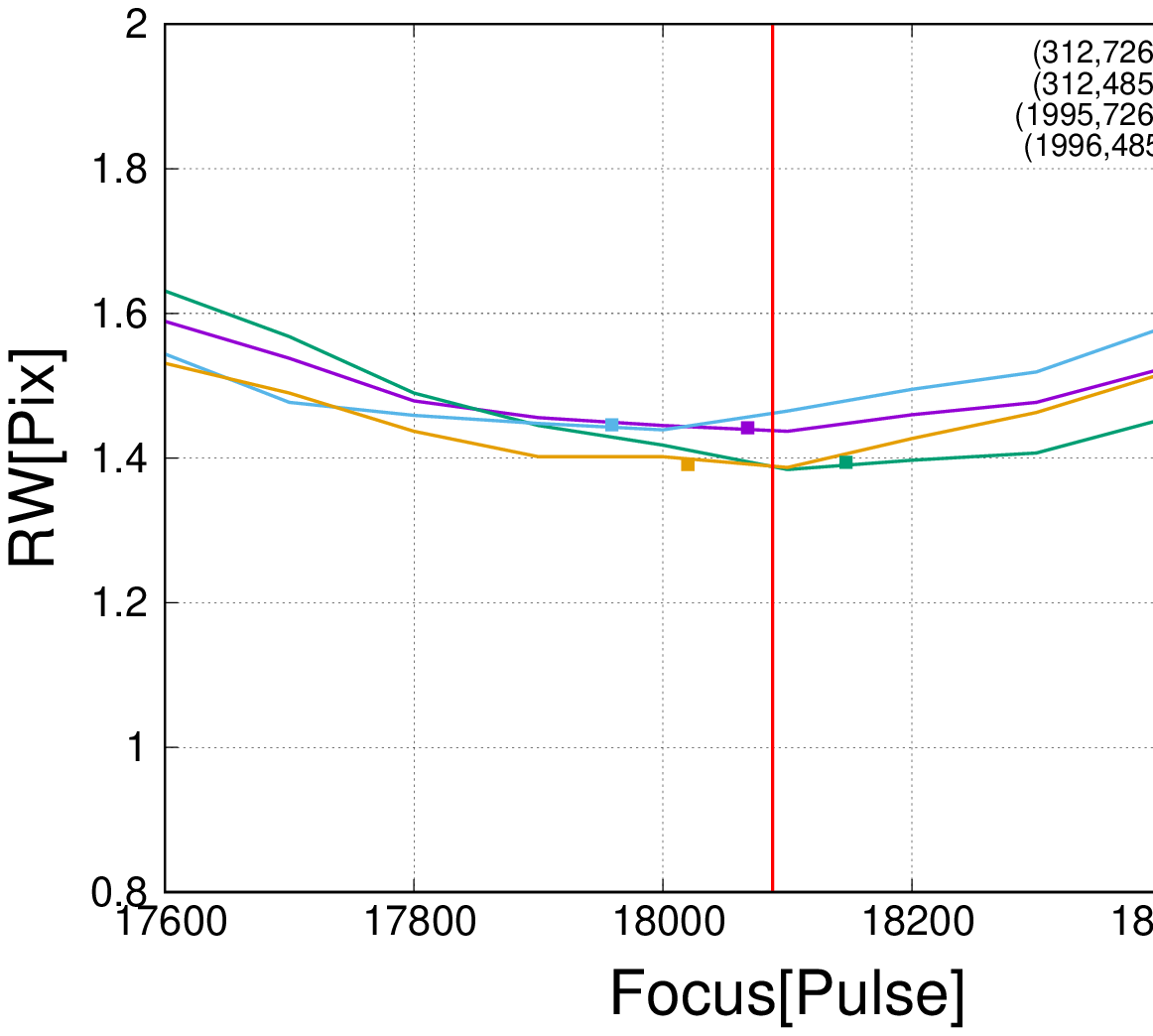}
  \includegraphics[width=8cm]{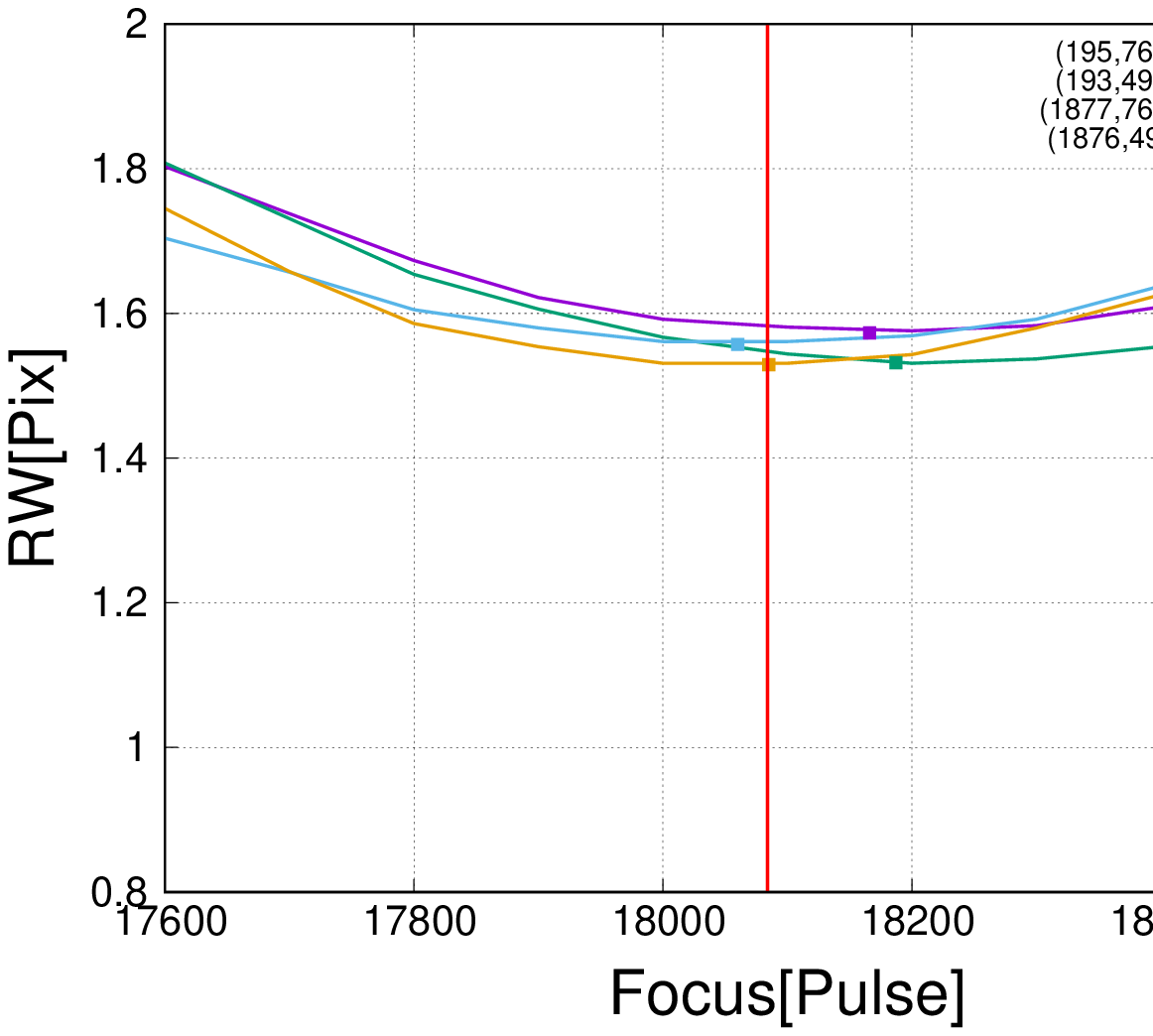}
 \end{center}
 \caption{Focus curves showing the spot size variation as a function of slit position displacement for four reference fiber emission line spots on the two detectors (upper panels: UV/Blue channel detector, lower panels: Red channel detector; see Figure~\ref{fig:anchor_images} for spot locations). Small squares indicate the minimum positions derived from quadratic polynomial fits to each curve, showing that all eight spots reach minimum size at nearly identical slit positions. The red vertical dashed lines represent the overall best-focus positions determined from each detector, accounting for the relative focal offset between them. The vertical axis denotes the square root of the second-order moment of the spot profile, corresponding to a $1\sigma$ image radius. Both detectors achieve an optimal spot size of approximately $3\text{ pixels}$, perfectly consistent with the optical design.
 {Alt text: A plot showing two sets of four closely overlapping curves, where the minimum points for all eight curves are well-aligned near the same slit position axis.}
}\label{fig:focus_curves}
\end{figure}

Prior to commencing any scientific observations with MIDSSAR, the focus alignment of the spectrograph must be executed. This procedure is performed by injecting emission-line light from the Fe-Ne hollow cathode lamp into the two reference fibers positioned at both longitudinal ends of the linear fiber slit. Consequently, the iron (Fe) lines are recorded in the UV/Blue channel, while the neon (Ne) lines are captured in the Red channel. To determine the optimal focus, the spatial and spectral extents of two pairs of bright emission lines located near the wavelength boundaries (yielding four monitoring positions per camera) are measured while systematically translating the fiber slit along the optical axis. This process generates focus curves (Figure~\ref{fig:focus_curves}) showing the spot size variation as a function of the slit position across the four locations on each detector. The final operational position of the fiber slit is defined where all monitored spot sizes are simultaneously minimized. Because the tilt of each detector and the relative focal positions between the two cameras have been carefully pre-aligned during the laboratory phase, these eight distinct spots reach their sharpest focus almost simultaneously.

\begin{figure}
 \begin{center}
  \includegraphics[width=8cm]{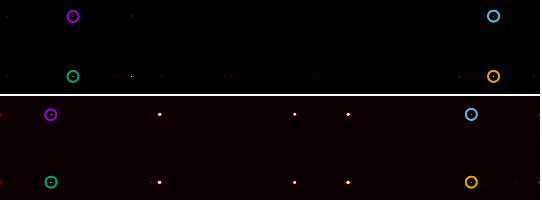}
 \end{center}
\caption{ Anchor images captured by the two cameras. The four circled emission spots correspond to the specific monitoring regions used during the focus alignment procedure. The pixel positions of these spots are subsequently referenced as master spatial and spectral coordinate origins to determine the precise locations of the entire scientific spectra on the detectors.
 {Alt text: Two detector images, each showing two parallel columns of emission-line spots arranged horizontally near the top and bottom of the frame, with four specific spots highlighted by circles.}
}\label{fig:anchor_images}
\end{figure}

Immediately following the focus alignment, the emission-line spectrum image of the reference fibers is secured for each camera. This frame serves as an ``anchor image'' (Figure~\ref{fig:anchor_images}) and is referenced throughout the subsequent data reduction pipeline to precisely track and identify the dynamic spatial and spectral positions of the individual fiber spectra on the detectors.

Laboratory thermal stability tests revealed that temperature variations induce a spectral position shift of approximately $1\text{ pixel }^\circ\text{C}^{-1}$, primarily along the spatial direction along the slit, with a characteristic thermal adaptation timescale of about 1 hour for the instrument.

In routine operations with the Seimei Telescope, the dome and its surrounding louvers are opened approximately 30 minutes prior to observations to allow the system to adapt to ambient temperatures. Since the Seimei Telescope employs a segmented primary mirror whose active control is sensitive to sharp thermal gradients, science observations are not executed during periods of extremely rapid temperature fluctuations. Once observations commence, the rate of temperature variation is generally reduced to at most $\sim 2^\circ\text{C }\text{h}^{-1}$. During the initial phase of observations when thermal drifts are relatively active, acquiring anchor images at $\sim 30$-minute intervals is sufficient to track spectral shifts. Once the instrument temperature stabilizes further, anchor frames need only be obtained every 1--2 hours, which is routinely executed during target transitions.

During data reduction, the exact spectral position for a science frame is estimated by time-interpolating between the two anchor frames taken before and after the science exposure. Additionally, the Quick Look (QL) pipeline continuously tracks spectral positions during observations and displays an alert prompting the observer to take a new anchor image whenever a shift of 1 pixel is detected.

\subsubsection{LED Flat Frames}\label{ssec:312}

\begin{figure}
 \begin{center}
  \includegraphics[width=8cm]{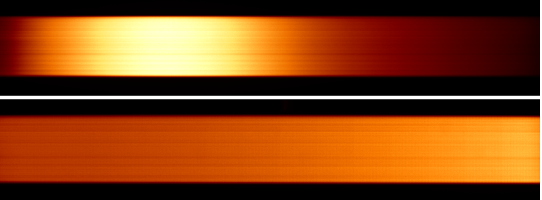}
 \end{center}
\caption{Defocused LED continuum spectrum images used for flat-fielding. The upper panel displays the UV/Blue channel image capturing the emission spectrum from the UV LED, while the lower panel displays the Red channel image capturing the spectrum from the white LED.
 {Alt text: Two detector images showing broad, belt-like continuum spectra horizontally stretching across each frame.} 
}\label{fig:out_of_focus_flats}
\end{figure}

Because the physical positions of the spectra projected onto the MIDSSAR detectors undergo gradual drifts due to ambient temperature variations, a standard continuum spectrum frame captured at the optimal focus position quickly becomes obsolete as a calibration master flat field over time. To mitigate this geometric instability, MIDSSAR employs a unique out-of-focus flat-fielding technique. The fiber slit is intentionally translated by $\pm 2\text{ mm}$ along the optical axis relative to the best focus position to secure two sets of out-of-focus continuum images (Figure~\ref{fig:out_of_focus_flats}). A total of 100 frames are acquired at each of the two defocused positions, and their combined average is utilized for the subsequent flat-field analysis. This substantial slit displacement blurs the point-source profile into a large spot with a diameter of approximately $25\text{ pixels}$ on the detector. Within this blurred patch, local pixel-to-pixel relative quantum efficiency variations can be effectively smoothed and calibrated.

The reduction pipeline processes these defocused continuum frames through a specialized multi-step algorithm to isolate the high-frequency pixel sensitivity variations:
\begin{enumerate}
    \item \textbf{Dark Subtraction and Periodic 4$\times$4 Flat Generation:} After subtracting an equivalent number of dark frames from the combined out-of-focus image, the frame is tile-segmented into small $4\times4$ pixel patches. This specific grid size corresponds to the sub-regions read out during a single A/D conversion cycle, which governs the primary periodic gain variations characteristic of these detectors. Each $4\times4$ patch is normalized to a mean value of unity and then stacked together to construct a master periodic $4\times4$ flat matrix. Replicating this matrix periodically across the entire detector area effectively eliminates the coarse, sensor-dependent electronic gain structures.
    \item \textbf{Removal of Spectral and Fiber Profiles:} To eliminate the intrinsic wavelength characteristics of the LED light source, the entire image is divided by a master one-dimensional profile obtained by integrating the dataset along the spatial slit direction. Subsequently, the image is normalized line by line so that the mean value of each horizontal line equals unity, thereby effectively neutralizing the variations in throughput among individual fibers (albeit spatially blended due to the intentional defocusing).
    \item \textbf{Large-scale Gradient Fitting:} The remaining global, large-scale spatial gradients are modeled by fitting a low-order surface---specifically utilizing a fourth-order polynomial along the wavelength (dispersion) axis and a linear function along the spatial slit axis. Dividing the image by this smooth surface removes the large-scale variations.
    \item \textbf{Synthesis of High-frequency Flat:} The residual component after these steps represents the high-spatial-frequency flat-field signatures. Finally, the previously isolated periodic $4\times4$ matrix component is multiplied back into this high-frequency residual frame. This process yields the final ``high-frequency flat'' master frame, which is optimized to precisely correct for localized, small-scale pixel sensitivity variations.
\end{enumerate}

\subsubsection{Wavelength Calibration and Sky Flat Frames}\label{sssec:313}

\begin{figure}
 \begin{center}
  \includegraphics[width=8cm]{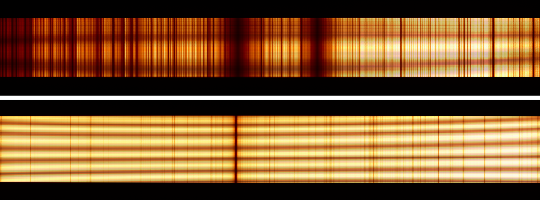}
 \end{center}
 \caption{Sky spectrum acquired with MIDSSAR. The prominent Ca$\,\text{II}$ H\&K absorption lines are clearly visible in the UV/Blue channel, while the $\text{H}\alpha$ absorption line is detected in the Red channel.
 {Alt text: Two detector images stacked vertically, each displaying 61 parallel, vertically aligned solar spectra with noticeable absorption features.}
 }\label{fig:sky_spectrum}
\end{figure}

Wavelength calibration begins by identifying the prominent emission lines recorded from the reference fibers. In the UV/Blue channel, the observed iron (Fe) lines are cross-matched with the atomic line database provided by the National Institute of Standards and Technology (NIST),\footnote{https://physics.nist.gov/PhysRefData/ASD/lines\_form.html} whereas the neon (Ne) lines captured in the Red channel are identified using the corresponding NIST neon line database. Because the internal Fe-Ne calibration lamp light can only be injected into the dedicated reference fibers, the primary wavelength calibration for the remaining scientific fiber slit array is conducted using daylight sky spectra---specifically, scattered sunlight (Figure~\ref{fig:sky_spectrum}). 

\begin{figure}
 \begin{center}
  \includegraphics[width=4cm]{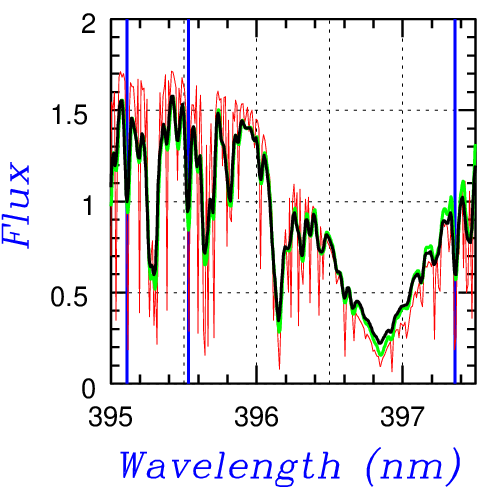}
  \includegraphics[width=4cm]{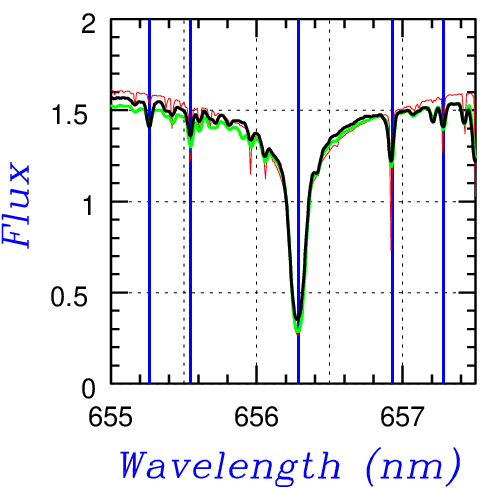}
 \end{center}
\caption{Comparison of the acquired sky spectrum with the solar model spectrum and the smoothed average NEID solar spectrum. The thick black lines represent the MIDSSAR sky spectrum wavelength-calibrated using the Fe-Ne lamp, the thin lines show the solar model spectrum from BASS2000, and the light thick lines denote the smoothed average NEID solar spectrum. Vertical lines indicate the positions of the absorption lines selected for wavelength calibration. The close-up panels show the profiles around the Ca$\,\text{II}$~H and $\text{H}\alpha$ absorption features.
{Alt text: A spectral plot with wavelength on the horizontal axis and relative intensity on the vertical axis, showing three highly overlapping profiles representing the observed data, the model spectrum, and the smoothed NEID solar spectrum.}
}\label{fig:solar_comparison}
\end{figure}

To verify the applicability of solar absorption lines as an independent wavelength reference, we utilized high-resolution solar spectra from the NEID spectrograph public archive\footnote{https://neid.ipac.caltech.edu/search\_solar.php}. Approximately 37,000 solar spectra acquired between February and July 2026 were downloaded and smoothed to match the spectral resolution of MIDSSAR, yielding a smoothed average NEID solar spectrum. A central wavelength comparison between the Fe-Ne arc lamp calibration and the smoothed average NEID solar spectrum demonstrated that the two independent calibration methods agree well within an error of approximately $0.2\text{ pixels}$. During this analysis, however, our temporal tracking revealed that certain solar absorption lines exhibit long-term wavelength shifts corresponding to up to $\sim 0.5\text{ pixels}$. Excluding these unstable features, we carefully selected a final set of 66 absorption lines for the UV/Blue channel and 43 lines for the Red channel to serve as stable wavelength reference anchors. Figure~\ref{fig:solar_comparison} shows the MIDSSAR sky spectrum compared with the smoothed average NEID solar spectrum alongside the BASS2000 solar model spectrum\footnote{https://bass2000.obspm.fr/solar\_spect.php}. Excellent agreement among these spectra is clearly confirmed.

To maintain this framework, a continuous sequence of 100 sky spectrum frames is observed once every few weeks. This periodic calibration run not only updates the core wavelength solution, but also simultaneously characterizes the relative wavelength response profiles across all 61 science fibers (spectral flat) and determines their relative throughput uniformities (fiber flat). During this sky spectrum acquisition process, the pipeline simultaneously records the precise pixel coordinates where the reference fiber emission lines ought to be located. This coordinate set is subsequently defined as the absolute master anchor position, serving as the benchmark reference for tracking thermal drifts in all subsequent science observations.

\subsection{Image Analysis}\label{ssec:32}

\subsubsection{Bias and Flat-Field Processing}\label{sssec:321}

Because the primary observation targets for MIDSSAR are relatively bright stars with magnitudes around 10th mag, maintaining a high signal-to-noise ratio ($S/N$) throughout the data reduction pipeline is paramount. Consequently, rather than subtracting standard dark frames with matching exposure times---which are typically acquired at a different time from the science observations and can introduce unnecessary statistical Poisson noise to bright-target data---the reduction process begins by subtracting a master bias frame constructed by averaging over 100 bias images. Following the bias subtraction, pixel regions affected by sharp, spike-like cosmic-ray noise are flagged and replaced with $\text{NaN}$ values. The science frame is then divided by the pre-generated high-frequency flat master frame (described in Section~\ref{ssec:312}) to execute localized, small-scale pixel sensitivity corrections.

Crucially, on this specific detector, the dominant dark component does not originate from standard thermal dark current but is instead caused by a faint infrared light emission known as amp-glow, which is continuously radiated from the peripheral readout circuitry during exposures. Because this amp-glow is manifested as actual photons incident on the sensor, it undergoes the exact same pixel-to-pixel sensitivity modulations as the incoming scientific starlight. Therefore, the pipeline is uniquely structured to apply the high-frequency flat-field division \textit{prior} to the dark background subtraction. 

\begin{figure}
 \begin{center}
  \includegraphics[width=8cm]{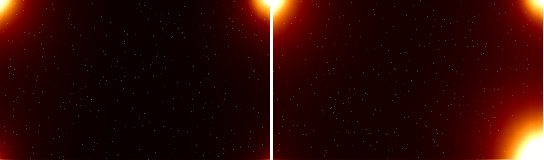}
 \end{center}
 \caption{Dark template images for the (left) UV/Blue channel and (right) Red channel. The parasitic background is dominated by the amp-glow component originating from the peripheral readout circuitry. Because the intensity ratios and spatial patterns of this emission vary from day to day, a combination of several distinct pre-compiled templates is utilized to dynamically synthesize and subtract the optimal dark baseline.
 {Alt text: Two detector images placed side by side, where several of the four corners are noticeably glowing due to structured amp-glow artifacts.}
 }\label{fig:template_dark}
\end{figure}

This photon-induced amp-glow component contributes a parasitic signal of approximately $0.02\text{--}0.04\text{ e}^-\,\text{ s}^{-1}\text{ pixel}^{-1}$ near the central region of the detector, while becoming considerably brighter near the outer edges. Fortunately, its impact on the scientific data remains minimal because the fiber spectra are strictly projected onto the central region. To precisely eliminate this structured artifact, several distinct template dark images (Figure~\ref{fig:template_dark}) are prepared in advance by spending a substantial amount of calibration time on multiple, deep $2000\text{-second}$ dark exposures. During the execution of the reduction pipeline, the software simply references these pre-compiled archival templates; the optimal background model is then dynamically synthesized via a linear combination of these templates to find the best-fit intensity distribution profile, which is subsequently subtracted from the flat-fielded science frame.

\subsubsection{Image Transformation and Advanced Data Processing}\label{sssec:322}

The raw spectra projected onto the detector are aligned such that the wavelength dispersion axis is approximately horizontal and the spatial slit axis is approximately vertical; however, they inherently suffer from subtle distortions induced by slight sensor rotation and optical distortion aberrations. In general astronomical data reduction, geometric transformations like rotation or resampling are highly discouraged because they inevitably blend informational content among neighboring pixels. Nevertheless, the MIDSSAR pipeline overcomes this degradation by strictly permitting only a single image transformation session executed through the following sophisticated nine-step procedure:

\begin{enumerate}
\item \textbf{Pixel Subdivision:} Rather than isolating individual fiber traces, the pipeline crops a single, continuous belt-like sub-frame encompassing all 61 spectral traces simultaneously. Specifically, a $2160 \times 360\text{ pixels}$ region containing the entire spectral array is extracted from the full $2160 \times 1280\text{ pixels}$ detector area. This cropped region is then upscaled to a high-resolution $4320 \times 720\text{ pixels}$ grid by subdividing the pixel size by half along both axes. In this subdivision process, the center coordinates of the original pixels and the subdivided sub-pixels are aligned, establishing new sub-pixel boundaries directly over the original pixel margins (Figure~\ref{fig:pixel_subdivision}). Since the Point Spread Function (PSF) and spectral intensity distribution naturally spread over several pixels, placing intermediate interpolated subpixel values along neighboring boundaries reproduces a spatial distribution that closer resembles the underlying continuous profile. This pixel subdivision process achieves three primary objectives:
\begin{itemize}
    \item The individual fiber spectra, originally spaced at intervals of $\sim 4.5\text{ pixels}$, become spaced at 9 subpixels after subdivision, allowing for standardized spatial handling across the detector array.
    \item It improves the precision of tracing spectral peak positions within small sampling windows in both the spatial and wavelength directions.
    \item It suppresses and confines the spatial blending of signal information during the subsequent image transformation (Step 3) within a radius of half the original pixel size.
\end{itemize}

\begin{figure}
 \begin{center}
  \includegraphics[width=8cm]{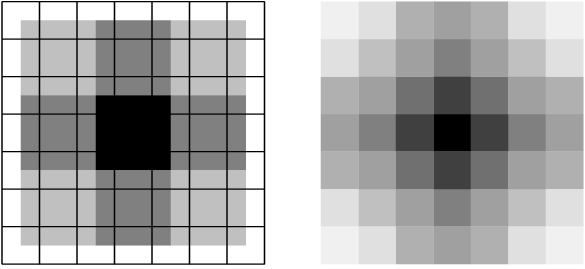}
 \end{center}
 \caption{Schematic diagram of the pixel subdivision process. For images where the point spread function spreads over several pixels, data values are expected to vary smoothly across adjacent pixels. Therefore, placing subpixels with intermediate values directly on the boundaries between neighboring pixels reproduces a distribution shape closer to the intrinsic physical profile.
 {Alt text: Two side-by-side grid diagrams demonstrating pixel subdivision, showing a brightness distribution transitioning from a 3 by 3 grid to a finer 7 by 7 grid.}
 }\label{fig:pixel_subdivision}
\end{figure}
    \item \textbf{Residual Background Subtraction:} The background signal levels measured in the unilluminated regions immediately above and below the spectral traces are treated as the residual dark artifacts that were not fully eliminated during the primary dark subtraction in Section~\ref{sssec:321}. These remaining background offsets are modeled across the spectral region via a vertical linear interpolation and subsequently subtracted to ensure a perfectly zero-centered baseline.
    \item \textbf{Geometric Transformation:} Using a distortion transformation vector map derived from the sky flat analysis, the sub-divided image undergoes a single geometric transformation. Through this process, the dispersion axis is perfectly rectified onto the horizontal axis---achieving a linear dispersion of $0.005\text{ nm}\text{ pixel}^{-1}$ for the UV/Blue channel and $0.006\text{ nm}\text{ pixel}^{-1}$ for the Red channel---while the spatial slit axis is re-aligned vertically, ensuring that the individual fiber centers are uniformly spaced at intervals of $9\text{ pixels}$.
    \item \textbf{Spectral Flat Correction:} The extracted spectrum of each individual fiber is divided by the pre-computed spectral flat profile (described in Section~\ref{sssec:313}) to seamlessly neutralize the intrinsic spectral throughput variations among the fibers.
    \item \textbf{Dynamic Template Synthesis:} Based on the profiles of the 61 science fibers, the pipeline classifies the data into high-count and low-count fiber cohorts to construct two separate, high-$S/N$ template spectra. The spectrum of each individual fiber is then modeled and fitted via a linear combination of these two templates.
    \item \textbf{Cosmic-Ray Flagging and Masking:} Pixels exhibiting large statistical residuals from the template fitting are flagged as transient noise artifacts, such as cosmic-ray strikes, and are replaced with $\text{NaN}$ values. Combining these newly flagged regions with the primary $\text{NaN}$ masks established in Section~\ref{sssec:321}, all $\text{NaN}$ pixels and their immediate surroundings are replaced with the local median count levels evaluated from the outer uncorrupted region.
    \item \textbf{Profile Reconstruction:} The smooth template fitting components subtracted in Step 5 are added back to the masked data to fully restore the intrinsic stellar spectral features.
    \item \textbf{Iterative Refinement:} Steps 5 through 7 are iteratively executed for three full cycles to achieve numerical convergence in cosmic-ray rejection and profile preservation.
    \item \textbf{Final Extraction and Spatial Re-synthesis:} The 61 individual fiber profiles are combined via a weighted integration, using the mean count of each fiber as the weighting factor, to generate and export the final optimized one-dimensional spectrum. Simultaneously, the mean count of each fiber is divided by the fiber flat and mapped onto the original bundle layout utilizing the known physical correspondence between the slit position and the hexagonal fiber bundle arrangement. This step outputs a re-synthesized spatial image of the target object on the bundle face.
\end{enumerate}

This unified reduction flow successfully extracts and consolidates the raw detector imagery into two core components, which is equivalent to decomposing three-dimensional (3D) spectroscopic data into its constituent spectral and spatial dimensions. Specifically, it yields the final optimized one-dimensional stellar spectrum along the wavelength axis and the re-synthesized two-dimensional image shape of the astronomical target on the fiber bundle surface. 

The transformation vector map applied in Step 3 is explicitly calculated to rectify the sky flat frame under stringent optimization constraints. Because higher-order optical aberrations remain extremely stable over time, the vector map does not change drastically, except for pure translational shifts driven by positional drift or magnification scaling induced by focal length variations.

To actively correct for dynamically evolving thermal deformations during science observations, the pipeline dynamically tracks the system behavior using reference anchor images. Specifically, the pipeline evaluates the closest anchoring frames taken before and after each science observation, applying a time-linear interpolation between them to estimate the exact anchor coordinates at the midpoint of the science exposure. A primary affine matrix is then computed to transform these estimated coordinates back into the master anchor positions defined during the sky flat acquisition. This dynamic correction vector is linearly added to the master distortion vector map prior to resampling, ensuring that any subtle image shifts caused by internal thermal drifts are perfectly neutralized.

Throughout the entire workflow---starting from the primary processing in Section~\ref{sssec:321} up to this final stage---the pipeline continuously tracks and error-propagates every single photon count on a pixel-by-pixel basis to enable rigorous noise estimation. By combining this integrated count log with the electronics read gain factor, the exact variance and localized noise levels are determined. Concurrently, the pipeline automatically calculates the equivalent widths ($EW$) of the key spectral features against the local continuum level across the $\text{Ca}\,\text{II}$ H\&K and $\text{H}\alpha$ line regions, utilizing two distinct bandpass extraction windows with widths of $\pm 30\text{ pixels}$ and $\pm 300\text{ pixels}$.

\begin{figure}
 \begin{center}
  \includegraphics[width=8cm]{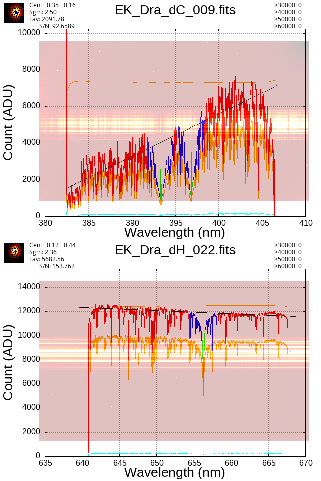}
 \end{center}
 \caption{Integrated data reduction summary and diagnostic display generated by the pipeline. The spectral bandpasses utilized for the automatic equivalent width calculations are plotted in distinctive colors, with the estimated continuum levels indicated by solid straight lines.To evaluate the impact of background light on the extracted spectra, the profile is co-plotted with both the absolute spectrum obtained after subtracting the average of the low-count peripheral regions and the ratio of this subtracted spectrum to the profile before subtraction. In the upper-left panel, the peak centroid position of the re-synthesized target image is highlighted in blue. For further details on the other metrics, see the main text.
 {Alt text: For each of the two channels, the comprehensive diagnostic panel overlays the final astronomical spectrum against a faintly blended raw detector background, with a reconstructed target morphology display in the upper-left quadrant along with various quantitative observational metrics derived from the data image.}
 }\label{fig:diagnostic_summary}
\end{figure}

Ultimately, all of these processed parameters, diagnostics, and images are consolidated and visualized within a single summary frame, as shown in Figure~\ref{fig:diagnostic_summary}. This integrated diagnostic display overlays the extracted final one-dimensional spectrum, the estimated noise level curve, and the reconstructed spatial target shape on the fiber bundle, all layered over the raw detector imagery displayed as a faint background image. To facilitate real-time performance tracking, the upper-left corner of the frame prominently displays key observational metrics: the centroid position of the stellar target, the $2\sigma$ spatial image size expressed in units of arcseconds, the average counts recorded across the fiber array, and the average $S/N$ achieved across the entire wavelength range. Meanwhile, the upper-right corner monitors detector saturation risks by reporting the exact number of pixels exceeding threshold count levels of $30000$, $40000$, $50000$, and $60000\text{ counts}$.

\subsubsection{Quick Look Processing}\label{sssec:323}

During real-time observations, the data reduction pipeline operates entirely autonomously without requiring manual command inputs. It actively monitors the storage directory and triggers the analysis sequence the instant a newly acquired science image is saved. However, while a single frame processing cycle is actively underway, the pipeline completely suspends its file-checking routine. Upon the successful completion of the current processing cycle, the software then checks the directory for any newly generated frames. If multiple science images have accumulated during that interval due to the execution time exceeding the short exposure cadence, the pipeline automatically bypasses all intermediate pending frames and exclusively targets the single most recently saved raw file for the next reduction cycle. The total processing latency for a single science frame is approximately $5\text{ s}$, after which the integrated diagnostic summary frame (described in Section~\ref{sssec:322} and Figure~\ref{fig:diagnostic_summary}) is immediately rendered and refreshed on the control web interface.

\begin{figure}
 \begin{center}
  \includegraphics[width=8cm]{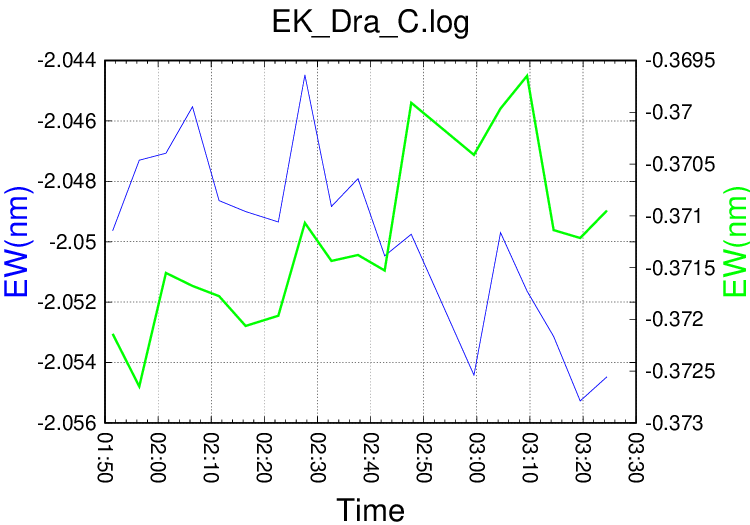}
  \includegraphics[width=8cm]{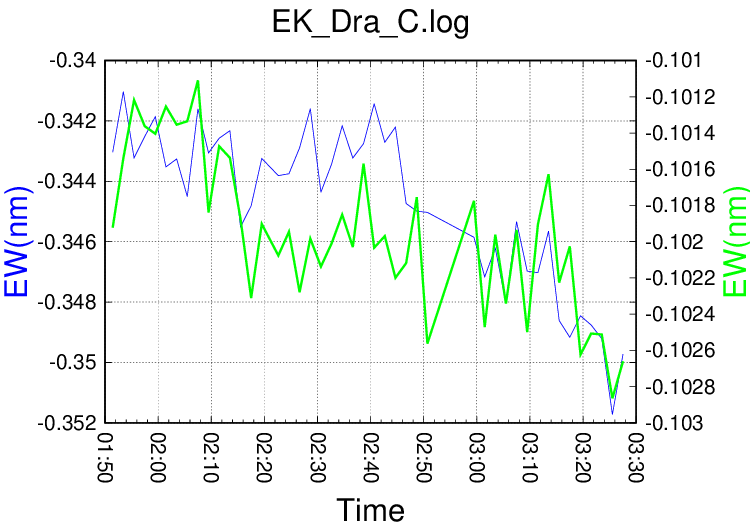}
 \end{center}
 \caption{Real-time equivalent width ($EW$) monitoring panel displayed on the Quick Look web interface. Thick lines represent the $EW$ integrated within the $\pm 30\text{ pixels}$ bandpass window, while thin lines represent the $\pm 300\text{ pixels}$ window, matching the corresponding color-coded features shown in Figure~\ref{fig:diagnostic_summary}. The plotting interface is dynamically refreshed in real time by continuously appending the latest calculated metrics as each new science frame is processed.
 {Alt text: Two real-time line charts plotting the temporal evolution of equivalent widths on the vertical axis against time on the horizontal axis, with each chart displaying two distinct line profiles representing different extraction widths.}
 }\label{fig:ew_quicklook}
\end{figure}

A primary scientific objective of this real-time pipeline is to enable immediate, on-site detection of transient stellar flare events. To accomplish this, the equivalent widths ($EW$) of the key chromospheric diagnostic---specifically the $\text{Ca}\,\text{II}$ H\&K and $\text{H}\alpha$ line---are continuously monitored in real time using the two distinct extraction bandpass windows. These calculated values are dynamically appended and visualized as real-time light-curve plots on the web interface, as shown in Figure~\ref{fig:ew_quicklook}. This instantaneous feedback loop empowers observers to make critical, time-sensitive tactical decisions during an observing run, such as determining whether to continue monitoring the current active target or to immediately pivot the telescope toward another programmed target star.

\subsection{Test Observations and Performance Verification}\label{ssec:33}

The relationship between the prism rotation angles of the ADC and the telescope altitude is analytically predetermined during the optical design phase. Therefore, the primary objective of the on-sky calibration is to accurately determine the operational origin points (zero-point offsets) for the two prisms. Prior to science observations, these origin points were calibrated using the following two-step on-sky procedure:
\begin{enumerate}
    \item A target near the zenith was observed to determine the relative origin between the two prisms. This ensures that the object image falls on the same bundle position in both channels without introducing artificial dispersion.
    \item A target at a low altitude ($\sim 20^\circ$) was observed to align the absolute origin of the combined dispersion vector in the direction opposite to the atmospheric dispersion vector.
\end{enumerate}

To evaluate the on-sky performance of the ADC without being affected by short-term image motion due to atmospheric turbulence, we selected representative science observation frames with integration times ranging from 1 to 10 minutes (over which atmospheric seeing is naturally time-averaged) rather than using short-exposure calibration frames.

\begin{figure}
 \begin{center}
  \includegraphics[width=8cm]{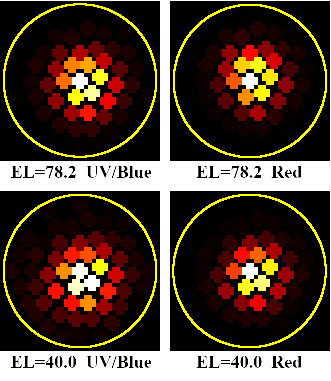}
 \end{center}
 \caption{Target images captured on the fiber bundle during actual science observations at different target altitudes. Upper panels: altitude angle of $78.2^\circ$ with integration times of $100\text{ s}$ (UV/Blue) and $60\text{ s}$ (Red). Lower panels: altitude angle of $40.0^\circ$ with integration times of $600\text{ s}$ (UV/Blue) and $300\text{ s}$ (Red). The circles in each panel indicate a region with a diameter of $5''$. The ADC successfully maintains the target near the center of the bundle regardless of the target altitude.
 {Alt text: Four object images on the fiber bundle are displayed side-by-side, each surrounded by a circle indicating the 5-arcsecond diameter field.}
 }\label{fig:adc_onsky}
\end{figure}

Figure~\ref{fig:adc_onsky} shows the target images on the optical fiber bundle captured during actual science observations at altitude angles of $78.2^\circ$ and $40.0^\circ$. As demonstrated in Figure~\ref{fig:adc_onsky}, the target centroid remains reliably positioned near the center of the bundle (with an image size of $\sim 2''$) in both the UV/Blue and Red channels regardless of the target altitude. This confirms that residual atmospheric dispersion is successfully minimized, ensuring stable coupling efficiency and preventing degradation of the final signal-to-noise ratio.

The primary requirement for any newly developed astronomical instrument is the rigorous stability of its observed spectra. It is critical to quantify the extent to which the intensities and profiles of emission/absorption lines, as well as the overall continuum slopes, are impacted by dynamic variations in the instrumental state, ambient weather conditions, and atmospheric seeing. To comprehensively evaluate these stability characteristics under realistic, non-ideal operational conditions, we executed the following verification procedures.

\begin{enumerate}
    \item We conducted a continuous $90\text{-min}$ time-series observation of the active solar analog EK Dra ($B \approx 8.2\text{ mag}$, $R \approx 7.3\text{ mag}$). The individual exposure times per frame were set to $5\text{ min}$ for the UV/Blue channel and $2\text{ min}$ for the Red channel. Over the course of this $90\text{-min}$ tracking window, the elevation angle of the target object continuously decreased from $45^\circ$ to $35^\circ$. Concurrently, the apparent brightness of the star temporarily degraded to two-thirds of its maximum value due to the passage of thin clouds, and the atmospheric seeing fluctuated by approximately 20\%. Figure~\ref{fig:ew_quicklook} corresponds to a representative single summary frame obtained during this specific observation run.
    \item Each extracted individual spectrum was divided by the grand-average spectrum integrated over the entire $90\text{-min}$ duration, and subsequently normalized so that the mean continuum level equals unity.
\end{enumerate}

\begin{figure}
 \begin{center}
  \includegraphics[width=8cm]{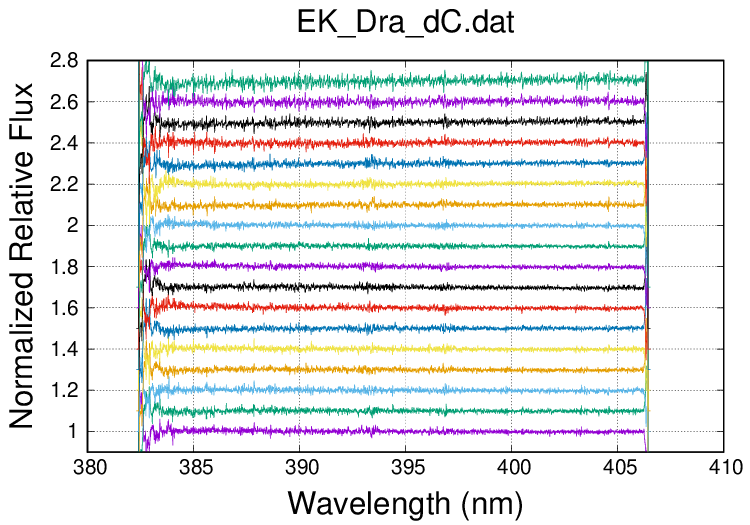}
  \includegraphics[width=8cm]{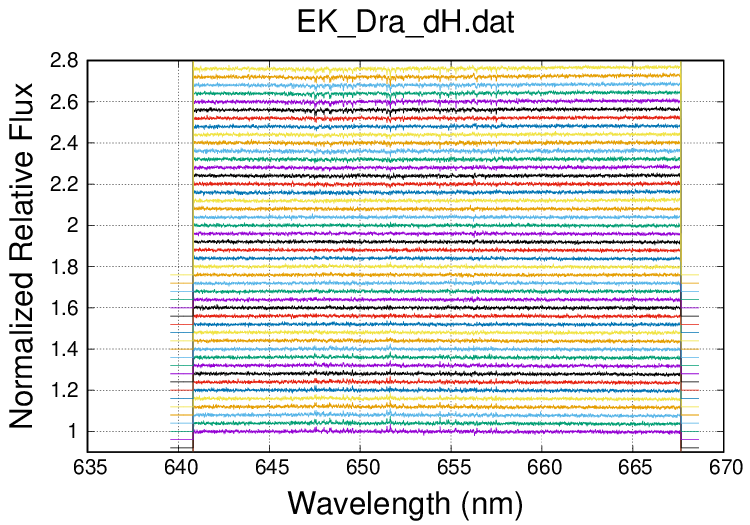}
 \end{center}
 \caption{Stability of the relative spectral profiles evaluated from the $90\text{-min}$ continuous monitoring of EK Dra. Individual spectra normalized by the time-averaged spectrum are vertically offset by increments of $0.1$ for the UV/Blue channel and $0.04$ for the Red channel to show the temporal variations clearly. The plot demonstrates that a highly stable spectral profile is maintained without being affected by various environmental and instrumental fluctuations.
 {Alt text: Two multi-panel plots showcasing the temporal stability of normalized spectral shapes against wavelength on the horizontal axis, where numerous nearly flat horizontal curves centered around unity are plotted with sequential vertical offsets for clarity.}
 }\label{fig:spectral_stability}
\end{figure}

The results of this stability analysis are displayed in Figure~\ref{fig:spectral_stability}. Although a minor intensity fluctuation in the continuum is recognizable at the shortest wavelength edge of the UV/Blue channel---which is highly likely attributed to the dynamic variations in atmospheric seeing---the remaining spectral regions exhibit sufficient stability against any environmental or instrumental changes. While EK Dra is a well-known active star capable of triggering energetic stellar flares, Figure~\ref{fig:ew_quicklook} clearly demonstrates that the equivalent widths ($EW$) of the chromospheric lines remained stable with negligible variations throughout this particular observation. Furthermore, from separate observations conducted under more severe meteorological fluctuations, we confirmed that when the total throughput counts drop below half of those under a clear sky due to thick cloud attenuation, the continuum of the Red channel experiences a slight redward tilting (reddening) effect.

\begin{figure}
 \begin{center}
  \includegraphics[width=8cm]{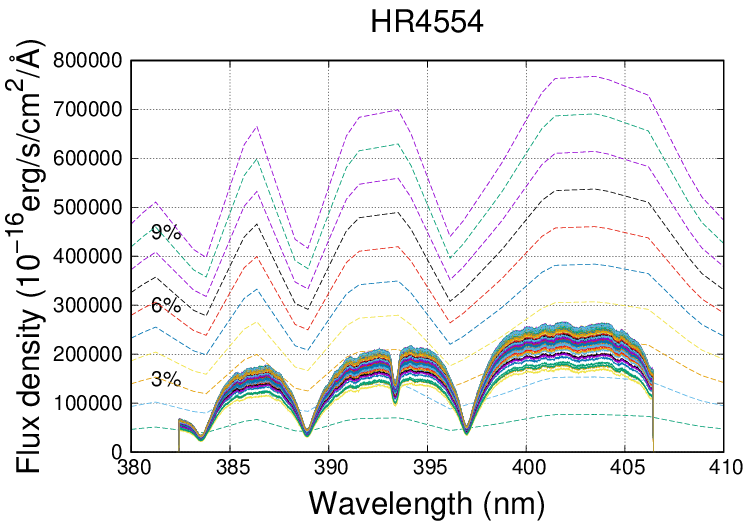}
  \includegraphics[width=8cm]{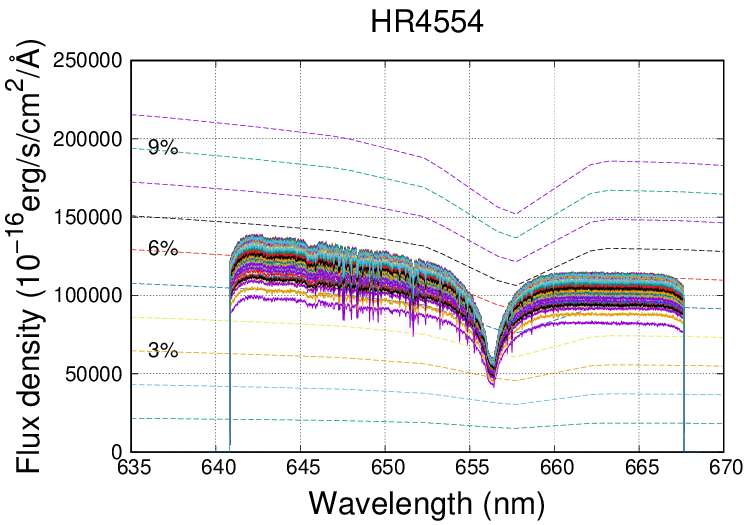}
 \end{center}
 \caption{Total throughput efficiency of the observation system. The observed data of HR4554 are converted to match the physical units of the spectrophotometric standard star catalog and are co-plotted within the same graphical interface. Dashed lines represent the reference catalog flux scaled by factors from 1\% to 10\% at increments of 1\%. Note that because the spectral resolution of the reference catalog is lower than that of MIDSSAR, the profiles around strong absorption lines do not perfectly match.
 {Alt text: Two line charts plotting the flux density on the vertical axis against wavelength on the horizontal axis, displaying the observed spectra co-plotted alongside multiple reference standard star spectra scaled by various percentage multipliers represented by dashed lines.}
 }\label{fig:system_efficiency}
\end{figure}

To evaluate the end-to-end throughput performance of the entire observation system, we conducted continuous observations of HR4554, which is selected from the ESO spectrophotometric standard stars catalog\footnote{https://www.eso.org/sci/observing/tools/standards/spectra/stanlis.html}, with an exposure time of $1\text{ s}$ per frame. The observed raw data were converted into physical flux units that allow for a direct, quantitative comparison with the reference catalog spectrum. The total system efficiency was subsequently derived by co-plotting and evaluating both the observed and cataloged profiles within the same graphical interface. As illustrated in Figure~\ref{fig:system_efficiency}, the derived total system efficiency---which comprehensively encompasses the transmission of the Earth's atmosphere, the telescope optics, and the internal components of the instrument---reaches a maximum peak of approximately 3.5\% in the UV/Blue channel and 6.7\% in the Red channel.

During the test observations, we targeted EK Dra ($B \approx 8.2\text{ mag}$, $R \approx 7.4\text{ mag}$) and CR Dra ($B \approx 10.9\text{ mag}$, $R \approx 9.0\text{ mag}$) under various meteorological conditions, which also served as a dual-purpose execution for flare monitoring. The exposure times were fixed at $300\text{ s}$ for the UV/Blue channel, and set to $90\text{ s}$ for CR Dra and $120\text{ s}$ for EK Dra in the Red channel. The elevation angles for both targets spanned a range of $35^\circ\text{--}60^\circ$. In the regime where the target counts are sufficiently large and the contributions from the detector readout noise and dark current are negligible, both the noise level and the resulting signal-to-noise ratio ($S/N$) theoretically scale with the square root of the target counts. 

\begin{figure}
 \begin{center}
  \includegraphics[width=8cm]{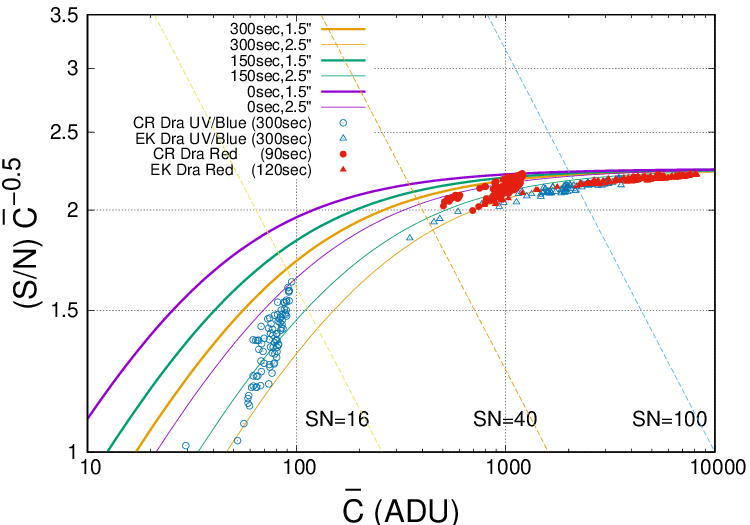}
 \end{center}
 \caption{Relationship between the average fiber count $\bar{C}$ and the $(S/N) \bar{C}^{-0.5}$ ratio. The circles and triangles represent the observation results for CR Dra and EK Dra, respectively, where open symbols denote the UV/Blue channel and filled symbols denote the Red channel. The solid curves indicate the theoretical values modeled under $1\farcs5$ seeing (thick lines) and $2\farcs5$ seeing (thin lines) conditions, displayed from left to right for exposure times of $\sim 0\text{ s}$, $150\text{ s}$, and $300\text{ s}$. Due to meteorological cloud attenuation, the observed data points are distributed leftward along these theoretical curves. The slanted dashed lines indicate the positions corresponding to $S/N = 16$, $40$, and $100$ from left to right.
 {Alt text: A scatter plot containing numerous observed data points categorized by various shapes alongside six theoretical limit curves and three vertical dashed lines, illustrating the precise correlation between the observed metrics and the modeled detection limits.}
 }\label{fig:sn_relation}
\end{figure}

Figure~\ref{fig:sn_relation} illustrates the relationship between the average count across the 61 fibers (denoted as $\bar{C}$) and the $S/N$ ratio divided by the square root of $\bar{C}$ (i.e., $(S/N) \bar{C}^{-0.5}$). The solid curves represent the theoretical values modeled assuming a conversion factor of $0.083\text{ e}^-/\text{ADU}$, a readout noise of $3.2\text{ e}^-\,\text{rms}/\text{pixel}$, an amplifier glow plus dark current component of $0.04\text{ e}^-/\text{s}$, and seeing sizes of $1\farcs5$ (enclosing 7 fibers) and $2\farcs5$ (enclosing 19 fibers).

When extracting individual fiber spectra from the subdivided image to construct the one-dimensional spectrum (Step 9 in Section~\ref{sssec:322}), an empirical average cross-sectional profile is first created directly from the science frame itself for each 9-subpixel-wide fiber trace. The spectral integration across the spatial profile is then executed using weights defined as the profile intensity divided by the variance (i.e., noise squared). Similarly, when co-adding the 61 individual fiber spectra, each fiber spectrum is weighted by its mean flux divided by its variance. This optimal weighting scheme ensures that the maximum attainable $S/N$ spectrum is consistently extracted, remaining robust against spatial trace shifts. Consequently, meteorological cloud attenuation manifests simply as a leftward shift of data points along the theoretical curves in Figure~\ref{fig:sn_relation}, whereas seeing variations correspond to transitions between the theoretical curves modeled for different seeing conditions.

Based on these empirical trends, achieving a benchmark of $S/N = 100$ requires an average count of $\bar{C} \approx 2000\text{ counts}$, which corresponds to a limiting magnitude of $B \approx 7.6\text{ mag}$ and $R \approx 8.1\text{ mag}$ for an exposure time of $1\text{ min}$. If the target star is $2\text{ mag}$ fainter, the expected $S/N$ degrades to slightly below $\sim 40$, a regime where the adverse impacts of the readout noise, amplifier glow, and dark current begin to visibly emerge. It should be noted that the photon counts and signal-to-noise ratios (S/N) utilized in this evaluation are based on the resampled values from the data reduction pipeline. During the pixel-subdivision process, the original counts from a single raw pixel are split into two pixels along the wavelength direction, combined with an additional expansion of a few percent required for exact wavelength allocation. Consequently, in the original raw images, the count values are a factor of two higher, and the S/N for bright targets scales up by a factor of $\sqrt{2}$. Accounting for these raw image characteristics, the aforementioned limiting magnitude values would increase by approximately $0.38\text{ mag}$.

\section{Conclusions}\label{sec:4}

We have successfully developed and commissioned MIDSSAR (MId Dispersion Spectrograph for Stellar Activity Research), a novel fiber-fed, dual-band spectrograph explicitly optimized for high-cadence monitoring of stellar superflares and active phenomena. Mounted on the 3.8-m Seimei Telescope at Kyoto University Okayama Observatory, MIDSSAR features a strategic operational integration with the existing high-dispersion spectrograph GAOES-RV, acting together as a single, multi-resolution spectroscopic system. By splitting the incoming light and utilizing specialized volume phase holographic and volume binary transmission gratings, the instrument achieves simultaneous, high-throughput observations of the critical chromospheric activity indicators: the Ca~{\sc ii} H\&K band ($382.5\mbox{--}406.5$~nm) at a resolving power of $R = 13,800$, and the H$\alpha$ band ($641.1\mbox{--}667.6$~nm) at $R = 18,200$. 

The quantitative performance and scientific readiness of the entire system have been successfully verified through laboratory calibrations and initial on-sky test observations conducted from March to May 2026. The key achievements demonstrated in this work are summarized as follows:
\begin{itemize}
    \item The geometric optical aberrations are strictly suppressed to within $1\text{ pixel}$ ($19\,\mu\text{m}$) across the CMOS detectors, ensuring that the instrumental profiles are thoroughly dominated by the physical fiber width. This provides the symmetry and stability in line profiles required for precise velocity-field measurements.
    \item The counter-rotating atmospheric dispersion corrector (ADC) successfully eliminates chromatic photon loss down to low elevation angles ($20^\circ$), enabling long-term, continuous monitoring tracking across large airmass variations.
    \item The robust control software and shared-memory architecture securely drive the two CMOS cameras of MIDSSAR during high-speed exposures down to $1\text{ ms}$ without data loss, while embedding real-time telescope telemetry.
    \item A dedicated data reduction pipeline has been fully established, capable of automatically correcting instrumental signatures, executing spatial reconstruction of the 61-fiber bundle profile, and extracting high-quality one-dimensional spectra with an optimized wavelength calibration precision of $\sim 0.002$~nm.
    \item On-sky evaluations confirmed excellent instrumental stability and system efficiency, yielding high-quality spectral time-series data. The resulting observed signal-to-noise ratio characteristics perfectly align with the theoretical curves modeled for different seeing conditions, confirming that the instrument successfully meets its primary design and scientific requirements as well as operational expectations.
\end{itemize}

With its stable long-term allocation on the 3.8-m Seimei Telescope and its unique capability to monitor the photosphere (via GAOES-RV) and the chromosphere (via MIDSSAR) simultaneously, this integrated instrument complex stands as a powerful new tool in stellar astrophysics. MIDSSAR is now fully operational and ready to execute systematic, high-cadence spectroscopic surveys of solar-type superflare stars and active low-mass dwarfs. The resulting data will be crucial for characterizing the energetic dynamics of stellar flares, estimating the mass and velocity scaling relations of stellar coronal mass ejections, and evaluating their profound impacts on exoplanetary space weather and habitability.

\begin{ack}
  Part of the data/observations used in this paper were obtained with the 3.8m Seimei Telescope operated by Kyoto University. We greedily appreciate the staff of Seimei Telescope for their support during the observations.
\end{ack}

\section*{Funding}
This work is supported by JSPS KAKENHI Grant Numbers JP24H00248(F.I., D.N., H.M., K.I., and K.S.), JP24K00680(F.I., D.N., H.M., and K.S.), JP24K00685(F.I, and H.M.), JP24K17082(K.I.), and JP25K01041(H.M.).

\end{document}